\documentclass[12pt]{article}
\usepackage{amsfonts}
\usepackage{bm, color}
\usepackage{amsmath}
\usepackage{epsfig}
\usepackage{hyperref}
\usepackage{breqn}
\usepackage{cancel}
\usepackage{cite}

\newcommand{\hu}{\hat u}
\newcommand{\hg}{\hat g}
\newcommand{\hT}{\hat T}

\newcommand{\hp}{\hat p}
\newcommand{\heps}{\hat \epsilon}
\newcommand{\uz}{\underline{z}}
\newcommand{\Tt}{T^{total}}
\newcommand{\cV}{{\cal V}}
\newcommand{\cD}{{\cal D}}

\newcommand{\cT}{{\cal T}}
\newcommand{\tot}{\text{total}}
\newcommand{\fluid}{\text{fluid}}
\newcommand{\pf}{\text{perfect}}
\newcommand{\diss}{\text{diss}}

\newcommand{\bea}{\begin{eqnarray}}
\newcommand{\eea}{\end{eqnarray}}
\newcommand{\rmd} {{\rm d}}
\newcommand{\fu}{{\mathfrak{u}}}

\renewcommand{\perp}{{P}}
\renewcommand{\hat}{\widehat}

\begin{document}

\pagestyle{plain}

\makeatletter
\@addtoreset{equation}{section}
\makeatother
\renewcommand{\theequation}{\thesection.\arabic{equation}}


\setcounter{page}{1} \setcounter{footnote}{0}


\begin{titlepage}


\begin{center}

\vskip 0cm

{\LARGE \bf  A New Approach
\\
to
\\
\vskip0.5cm
Non-Abelian Hydrodynamics
} \\[6mm]

\vskip 1cm

\textsc{Jose J. Fern\'andez-Melgarejo$^1$, Soo-Jong Rey$^{2,3}$,   \, Piotr Sur\'{o}wka$^{1, 4}$}\let\thefootnote\relax\footnote{josejuan@physics.harvard.edu, sjrey@snu.ac.kr, surowka@physics.harvard.edu}\\

\vskip 1cm

{\em
$^1$Center for the Fundamental Laws of Nature, Harvard University \\ Cambridge, MA 02138, \rm USA \\
\vskip 0.35cm
\em $^2$School of Physics \& Astronomy and Center for Theoretical Physics \\ Seoul National University, Seoul, 08826 \rm KOREA \\
\vskip 0.35cm
\em $^3$ B.W. Lee Center for Gauge, Gravity \& Strings \\ Institute for Basic Sciences, Daejeon, 34047 \rm KOREA\\
\vskip 0.35cm
\em $^4$Max-Planck-Institut f\"ur Physik (Werner-Heisenberg-Institut) \\ F\"ohringer Ring 6, D-80805 Munich, \rm GERMANY
}

\vskip 0.8cm

\end{center}

\vskip 1cm

\begin{center}

{\bf ABSTRACT}\\[3ex]

\begin{minipage}{14cm}
\small

We present a new approach to describe hydrodynamics carrying non-Abelian macroscopic degrees of freedom. Based on the Kaluza-Klein compactification of a higher-dimensional neutral dissipative fluid on a group manifold, we obtain a $d=4$ colored dissipative fluid coupled to Yang-Mills gauge field. We calculate the transport coefficients of the new fluid, which show the non-Abelian character of the gauge group. In particular, we obtain group-valued terms in the gradient expansions and response quantities such as the conductivity matrix and the chemical potentials. While using SU(2) for simplicity, this approach is applicable to any gauge group. Resulting a robust description of non-Abelian hydrodynamics, we discuss some links between this system and quark-gluon plasma and fluid/gravity duality.

\end{minipage}

\end{center}

\vfill

\end{titlepage}


\tableofcontents
\newpage
\rightline{\sl Denn die Menschen glauben an die Wahrheit dessen,}
\rightline{\sl was ersichtlich stark geglaubt wird.}
\rightline{\sl All truthful things are subject to interpretation.}
\rightline{\sl Which interpretation prevails at a given time}
\rightline{\sl  is a function of power, not truth.}
\rightline{-- Friedrich Nietzsche -- `The Will to Power'}
\vskip1cm

\section{Introduction}
\label{sec:intro}

Hydrodynamics has been an efficient approach for the description of strongly interacting state of matter. This boosted the research and application of hydrodynamics models, such as transport phenomena or hydrodynamic instabilities. One aspect in hydrodynamics that has not been explored in detail yet is the dynamics of a colored fluid charged under non-Abelian Yang-Mills gauge fields, where the constituents of the fluid carry non-Abelian color charges and interact with non-Abelian vectors. Due to its non-Abelian nature, we expect that this system gives rise to a variety of physical phenomena richer than its Abelian counterpart, viz. Maxwell plasma. Nevertheless, the level of rigor in formulating the theoretical foundations of this model and the understanding of its ensuing physical properties are far lesser.

A robust description will contribute to the characterization of some important physical systems. For example, the quark-gluon plasma behaves as an almost perfect dense fluid carrying color charge. However, the detailed microscopic understanding of the equilibration mechanisms after the heavy-ion collisions still happens to be an outstanding problem. A transient phase in the equilibration process is reached when the system is at local thermal equilibrium with yet non-equilibrated colored quark and gluon degrees of freedom (DOFs). Most of the analysis done so far is based on kinetic theory \cite{Heinz:1983nx,Heinz:1984yq,Mrowczynski:1987ch,Litim:2001db,Manuel:2003zr,Manuel:2006hg} and on the single-particle approach \cite{Wong1970}. Integrating out momentum, one obtains a covariant color continuity equation which, together with the mechanic conservation laws of the fluid, constitute the main equations of the system. Still, the construction of the required collision terms which enter the Boltzmann equation is highly non-trivial and except at weak coupling regime, there is no first-principles derivation. In addition, the applicability of kinetic theory is valid for not-so-far from equilibration situations. Consequently, we conclude that kinetic theory is a useful complementary tool but requires prior knowledge of the structure of the hydrodynamic equations.

Alternative approaches include the Poisson bracket formulation \cite{Holm:1984hg} and the action principle \cite{Jackiw:2000cd,Bistrovic:2002jx} of ideal fluid dynamics. Nevertheless, the study of dissipative effects, which constitute an integral part of hydrodynamics is well understood only at the level of the equations of motion (EOMs). The description of these effects at the level of an action requires placing the fluid on the Schwinger-Keldysh contour \cite{Grozdanov:2013dba}, which leads to auxiliary supersymmetric DOFs \cite{Haehl:2015uoc,Crossley:2015evo}.

Another aspect that shed light on the understanding of hydrodynamic structure is the duality between fluids and black holes \cite{Bhattacharyya:2008jc,Banerjee:2008th,Erdmenger:2008rm,Kanitscheider:2008kd}. This allowed us to discover previously neglected parity-breaking terms that were originated by quantum anomalies \cite{Son:2009tf,Landsteiner:2011cp,Loganayagam:2012pz}. To study non-Abelian DOFs coupled to fluids, we need a new background of black hole with non-Abelian Yang-Mills hair \cite{Torabian:2009qk,Eling:2010hu,Herzog:2014tpa}. However, in AdS/CFT correspondence, local symmetry in the bulk gravity is mapped to global symmetry in the boundary theory. Therefore, as the background field in the boundary theory is usually external and non-dynamical, we have no way of promoting non-Abelian global symmetries to gauge symmetries in the boundary theory.

For these reasons, we view this state of affair at odds: self-gravitating hydrodynamics, whose gravitational interaction is also intrinsically nonlinear, has been rigorously investigated in various contexts of relativistic astrophysics of compact objects \cite{Font:1998hf} and cosmology of large-scale structures \cite{Goroff:1986ep,Grinstein:1986sm}. We thus expect that non-Abelian hydrodynamics, at least at classical level, can also be rigorously formulated and investigated as much as the self-gravitating hydrodynamics. Such study would have a direct application to wider phenomena featuring non-Abelian DOFs such as the quark-gluon plasma \cite{ Jackiw:2004nm} and the spintronics with strong spin-orbit coupling \cite{RevModPhys.87.1213,Leurs2008}.

In this work we propose a completely new approach to bypass all the above conceptual and technical difficulties. We will consider a neutral and dissipative fluid coupled to Einstein gravity, in $D$ dimensions. We will do a Kaluza-Klein (KK) compactification \cite{Kaluza:1921tu,Klein:1926tv} of this system and obtain a fluid in $d=D-n$ dimensions whose constituents are charged under non-Abelian Yang-Mills fields, where $n$ is the dimension of the internal manifold. That is to say, we use KK dimensional reduction as a method to construct an \emph{ab initio} description of non-Abelian hydrodynamics. The KK compactification mechanism endows the lower-dimensional system with a set of gauge fields, the so-called KK gauge fields. The compactification ansatz and the topology of the internal manifold elucidate the resulting gauge symmetry of the $d$-dimensional system. As we are interested on non-Abelian hydrodynamics, we will do a compactification on an SU(2) group manifold \cite{Scherk:1979zr,Cho:1975sf}. Therefore, we will take $n=\text{dim}(G)=3$, where $G$ is the gauge group. We will perform this procedure on the EOMs of the starting fluid, which include dissipative terms\footnote{Heavy-ion collisions and other phenomenologically relevant phenomena occur off the equilibrium and consequently, dissipative effects result very important in their descriptions.}.

Our approach is based on the non-Abelian Kaluza-Klein compactification  on a SU(2) group manifold, which we interpret as an internal manifold whose isometry generates the non-Abelian color symmetry in the physical system. Since we start with a fluid from the outset, the resulting theory is in the long-wavelength expansion but will be coupled to new non-Abelian DOFs that the compactification generates.

\begin{figure}[ht!]
\vskip-1cm
\centering 
\includegraphics[angle=90, width=15cm]{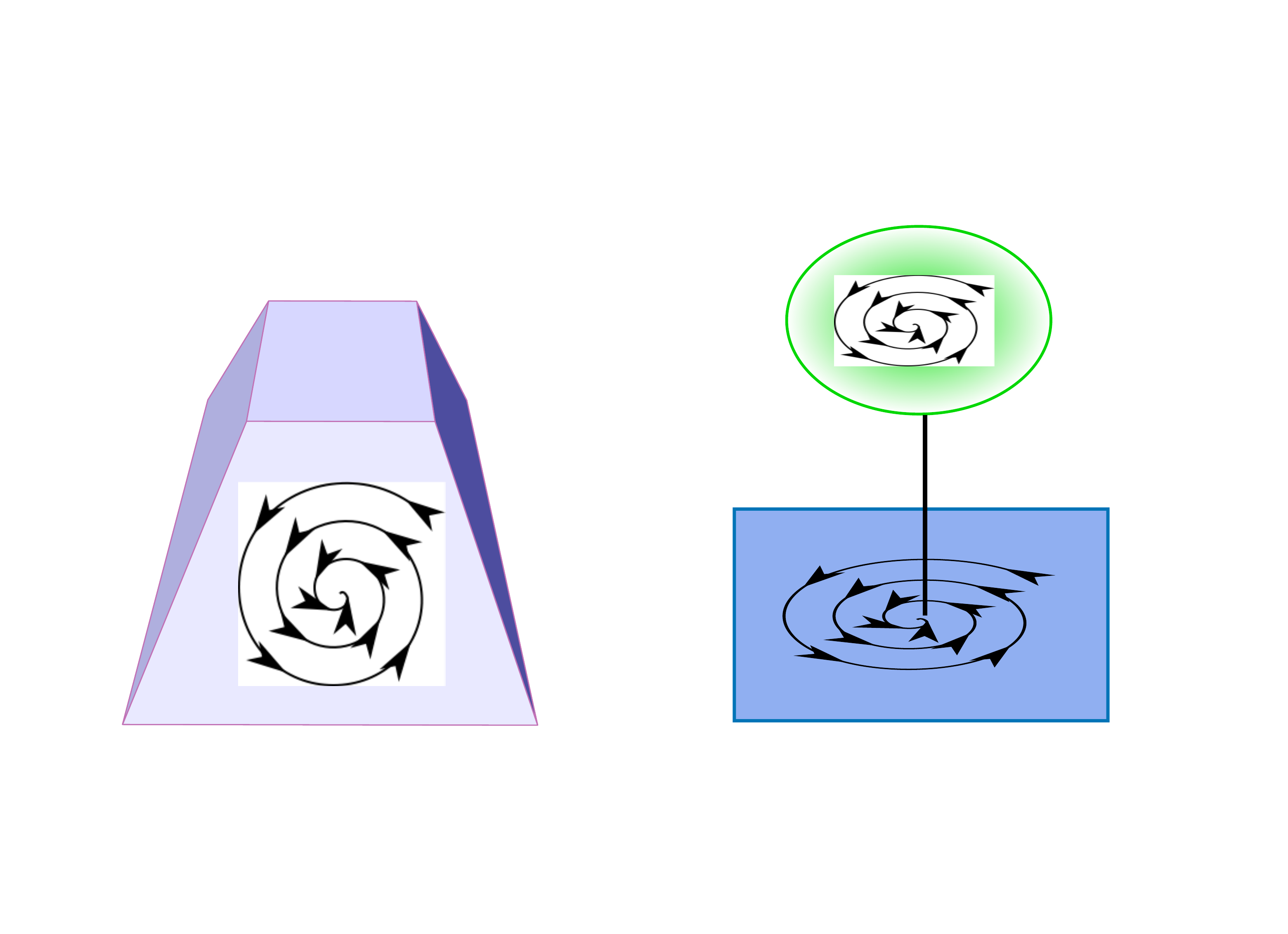}
\vskip-1cm
\caption{\sl  Our starting system is a $D$-dimensional dissipative fluid coupled to gravity (left). After KK compactification on a $n$-dimensional internal manifold with non-Abelian isometries, we obtain a $d$-dimensional dissipative fluid that, apart from being coupled to gravity, is charged under dynamic non-Abelian Yang-Mills gauge fields (right).}
\label{KKfigure}
\end{figure}
\vskip0.5cm

KK compactification provided a robust tool for the understanding of the (hidden) structure and the dynamics of gravity-matter systems, which descends from a more fundamental theory such as string/M-theories. If we start with a fundamental theory in $D$ dimensions defined on a manifold ${\cal M}_D$, we can find a stable solution of its equations of motion of the form ${\cal M}_D={\cal M}_d \times X_n$, where $d=(D-n)$, ${\cal M}_d$ is non-compact, reduced spacetime, and $X_n$ is a compact manifold of characteristic size $R$. At low energies, the compact space $X_n$ is not accessible by direct observations: it would take excitations of energy $E \sim 1/R$ to probe spacetime structures of a scale of order $R$. If $R$ is sufficiently small, this energy scale is gapped from the low-energy dynamics on ${\cal M}_d$. Nevertheless, the properties of $X_n$ will have important effects on the reduced theory. For example, if $X_n$ is a manifold with isometry group $G$, then metric fluctuations along the Killing directions of $X_n$ generate Yang-Mills gauge fields with gauge group $G$, which will be present in the dynamics of the lower-dimensional theory.

From the viewpoint of KK theory, a novelty of our work is that we include energy-momentum tensor of dissipative fluid, sourcing the Einstein field equations. The procedure, however, must be self-consistent. A KK compactification is said to be consistent if all the solutions of the $d$-dimensional theory satisfy the $D$-dimensional EOMs. In this work, we also present the necessary conditions to achieve a consistent reduction of fluid energy-momentum tensor.

Summarizing, the salient features of our approach are the following:
\begin{itemize}
\item The proposed KK method ``generates'' dynamical (non-)Abelian gauge fields which are self-consistently coupled to a charged/colored fluid.
\item This mechanism provides an {\sl ab initio} approach to (non-)Abelian hydrodynamics, distinct from gauge-gravity duality or fluid/gravity duality.
\item We apply the KK method to a neutral fluid at the outset coupled to gravity, thus bypassing kinetic theory.
\item The approach applies to dissipative fluids, for the compactification is at the level of equations of motion rather than action.
\end{itemize}

This paper is organized as follows. In Section 2 we present the main results of our work: they dynamics of the system, its symmetries and its properties. In the following sections we explain the KK dimensional reduction and the method to obtain our results. In particular, Section 3 reviews the basics of relativistic hydrodynamics and provides the necessary set-up and notations for our calculations. In Section 4, we review the dimensional reduction of the system Einstein-perfect fluid on a circle. This results in a fluid charged under a U(1) gauge field. In Section 5 we do the KK compactification on an SU(2) group manifold of the Einstein-dissipative fluid system  and study the conservation laws of the system. In Section 6 we evaluate our energy-momentum tensor and identify all the dissipative coefficients of the $d$-dimensional fluid.
In Section 7 we explain the main properties of our system and discuss future directions we are currently investigating. Appendices provide the details of our computations.


\section{Dissipative fluid dynamics with Yang-Mills charge}

In this section we will explain the dynamics and the main properties of a $d$-dimensional dissipative fluid that is charged under non-Abelian Yang-Mills gauge fields.

We denote space-time indices by $\mu,\nu,\rho=1,\ldots,d$, whereas indices $\alpha,\beta,\ldots= 1,\ldots,\text{dim}(G)$ correspond to the adjoint representation of the Yang-Mills group $G$. \footnote{$\text{dim}(G)$ will correspond to the dimension of the internal group manifold in the KK compactification.}

Let us consider the energy-momentum tensor associated to a dissipative fluid and non-Abelian Yang-Mills fields:
\begin{align}
\begin{split}
T^{\tot}_{\mu\nu}
= &
T^{\text{fluid}}_{\mu\nu}
+ \frac{1}{2} Q_c^{-2}(x)
\left(
	{\bf F}^\alpha{}_{\mu}{}^{\rho} {\bf F}^\alpha{}_{\rho \nu}
	- \frac12\eta_{\mu\nu} ({\bf F}^\gamma)^2
	\right)
\, ,
\end{split}
\end{align}
where ${\bf F}^\alpha{}_{\mu\nu}$ is the non-Abelian field strength of the gauge field ${\bf A}^\alpha{}_{\mu}$,  $Q_c$ is a coupling constant\footnote{From the KK perspective $Q_c=Q_c(x)$ is a scalar quantity that corresponds to the dilaton-dependent gauge coupling, $Q_c(x) \propto e^{\phi(x)}$. When we set $\phi=constant$, then $Q_c$ plays the role of a Yang-Mills coupling constant and it disappears from the EOMs of the gauge fields.} and
\begin{align}
T^{\text{fluid}}_{\mu\nu} = T^\pf_{\mu\nu} + T^{\rm diss}_{\mu\nu}
\, .
\end{align}
The perfect fluid piece is given by
\begin{align}
T^\pf_{\mu\nu}(x) = \left[\epsilon(x)+p(x)\right]
u_\mu 
u_\nu + p(x) \, \eta_{\mu\nu}
\, ,
\end{align}
where $p$ is pressure, $\epsilon$ is energy density, $u_\mu$ is the velocity field and $\eta_{\mu \nu}$ is Minkowski metric\footnote{Although in the following sections we will consider curved space-time, we can decouple gravitational interaction by taking the limit on the Newton's constant $G_N \rightarrow 0$ and assume flat space-time.}.

As for the dissipative part $T^{\rm diss}_{\mu\nu}$, in this description we will not choose any specific frame and will consider a generic energy-momentum tensor. Though further explicit assumptions are made in Section \ref{sec:su2diss}, we can generalize our results to any frame independent prescription.

The thermodynamic relation for the SU(2) charged perfect fluid that the we obtain after KK compactification accounts for the chemical potentials ${\boldsymbol \mu}^{\rm color}_\alpha$ associated to the color charges $\mathfrak Q_\alpha$,
\begin{align}
\epsilon+ p = \cT s + \mathfrak Q^\alpha {\boldsymbol \mu}^{\rm color}_\alpha
\, .
\end{align}
$\cT$ is the temperature and $s$ is the entropy density.

Let us specify  the dynamics of the system. The first EOM corresponds to the fluid dynamics evolution. Inspired by the KK compactification of a fluid coupled to gravity (in particular Biachi identities of Einstein equations,{\emph{cf.} Section \ref{sec:su2}), we obtain the conservation of the total energy-momentum tensor,
\begin{align}
\nabla^\mu T^{\tot}_{\mu\nu}
=
0
\, .
\end{align}
The second EOM describes the dynamics of the non-Abelian Yang-Mills gauge fields and introduces a non-Abelian colored current.
\begin{align}
\cD^\nu
{\bf F}^\alpha{}_{\mu\nu}(x)
=  
{\bf J}^\alpha_\mu (x)
\, .
\end{align}
The quantity ${\bf J}^{\rm color}_{\mu \, \alpha}(x)$ is a group-valued current that is covariantly conserved,
\begin{align}
{\bf J}^{\rm color}_{\mu \, \alpha}(x) =
Q_c \, \,
 {\mathfrak Q}_\alpha (x) \, u_\mu (x)
 + {\bf J}^{\rm diss}_{\mu \, \alpha}(x)
\, .
\end{align}
$\mathfrak Q_\alpha(x)$ is the color charge density attached to the fluid.

Although the explicit expression for the quantity ${\bf J}^{\rm diss}_{\mu \, \alpha}(x)$ in terms of the dissipative terms depends on the choice of the frame, the color current ${\bf J}^{\rm color}_{\mu \, \alpha}(x)$ will always be covariantly conserved. Further details for specific frame choices can be found in Section \ref{sec:su2diss}.

In our formulation, Lorentz force for a generic fluid $T^{\text{fluid}}_{\mu\nu}$ naturally emerges from the above energy-momentum tensor conservation law,
\begin{align}
\nabla^\nu T_{\nu\mu}^{\rm fluid}(x)   = Q_c \  {\bf F}^{\alpha \, \nu}{}_\mu(x) \ {\bf J}^{\rm color}_{\nu \, \alpha}(x)
\, .
\end{align}

Due to the frame-dependent structure of the energy-momentum tensor, departure from the Landau frame does not permit to read the transport coefficients associated with the dissipative terms from $T^\diss_{\mu\nu}$. To correctly identify these coefficients, we need a frame-invariant formulation of the dissipative terms which is in agreement with the 2nd law of thermodynamics,  $\nabla^\mu \mathfrak{J}^s{}_\mu \ge 0$, where $\mathfrak{J}^s{}_\mu$ is the entropy current. We propose the following generalization:
\begin{align}
\begin{split}
 \left(
	\perp_\mu{}^\rho(x) \perp_\nu{}^\lambda(x)
	- \frac{1}{d-n} \perp_{\mu\nu}(x) \perp^{\rho\lambda}(x)
	\right) T^\diss_{\rho\lambda}(x)
&= -2 \eta (x) \, \sigma_{\mu\nu}(x)
\, ,
\\
 \frac{\partial p(x)}{\partial \mathfrak Q_m} u^\mu(x) 
{\bf J}^\diss{}_{\mu m}(x)
+ \left(
	\frac{1}{d-n}\perp^{ab}
	-\frac{\partial p}{\partial \epsilon} u^\mu u^\nu
	\right)(x) T^\diss_{\mu\nu}(x)
&=
- \zeta (x) \theta (x)
\, ,
\\
\perp_\mu{}^{\nu}(x)\left(
	{\bf J}^\diss{}_{\nu m}
	+ \frac{\mathfrak Q_m}{\epsilon+p} u^\rho T^\diss_{\rho\nu}
	\right)
-
{\boldsymbol \kappa}_{mn}(x) \left(
	-\perp_\mu{}^\nu D_\nu\left( \frac{{\boldsymbol \mu}_n}{\cT}\right)
	+ \frac{1}{\cT}{\bf F}^n{}_{\mu\nu} u^\nu
	\right)
&=
0
\, ,
\end{split}
\end{align}
where $P_{\mu\nu}= \eta_{\mu\nu} + u_\mu u_\nu$ is a projector, $\boldsymbol{\kappa}_{mn}$ is the non-Abelian conductivity tensor, and $\eta$, $\zeta$, $\sigma_{\mu\nu}$ are the dissipative coefficients. Covariant derivatives are defined in Section \ref{sec:su2}.

This is the summary of the equations that govern our system. To establish this set of EOMs and conservation laws we have done the KK dimensional reduction of a higher-dimensional fluid. We have used the KK compactification as a guiding principle to obtain expressions that preserve SU(2) covariance and the conservation laws, which arise upon recasting the higher-dimensional ones.

In the following sections we will explicity show the calculations that lead to these equations.

\section{Kaluza-Klein approach}
\label{sec:hdfluid}

Our goal is to construct non-Abelian hydrodynamics. It consists of two components: colored fluid and Yang-Mills gauge field. Constructing its hydrodynamics starting from a microscopic Yang-Mills-matter theory (such as QCD) is just a theoretical idea: it is not feasible nor shedding light on physics. As such, we look for a mesoscopic approach. The idea is to utilize the Kaluza-Klein compactification to construct both components of non-Abelian hydrodynamics simultaneously. Our starting point is a self-gravitating, dissipative and neutral fluid in a dynamic $D$-dimensional spacetime ${\cal M}_D (\hat g_{MN})$, viz. a dissipative and neutral fluid coupled to the Einstein gravity, all in $D$ dimensions~\footnote{We  denote all $D$-dimensional variables as hatted quantities.}.  Our working assumption is that the $D$-dimensional matter is strongly interacting at the outset. While gravity is fundamentally weak, effective strength for the fluid depends on macroscopic conditions such as density and temperature.

\subsection{Self-gravitating dissipative fluid}
We will first characterize a neutral fluid in $D$-dimensional spacetime ${\cal M}_D(\hat g_{MN})$.  The hydrodynamic fields of the fluid consist of the velocity field $\hu ^M (\hat x)$ and other scalar quantities. The velocity field is normalized \footnote{We use the mostly plus signature.}
\begin{equation}\label{eq:uu1HD}
\hu^M (\hat x) \hu^N (\hat x) \, \hg_{M N} (\hat x) = -1
\, ,
\end{equation}
such that it carries $(D-1)$ independent components. On the other hand, the number of independent scalar quantities depends on the number of equations of state that we consider. As we are not considering any equation of state. For a perfect fluid, we will consider temperature $\hat T(\hat x)$, pressure $\hat p(\hat x)$ and energy density $\hat \epsilon(\hat x)$ to be independent quantities. The same applies for the dissipative coefficients: shear viscosity $\hat \eta$, bulk viscosity $\hat \zeta$, shear tensor $\hat \sigma_{AB}$ and expansion scalar $\hat \theta $ will be independent quantities of the $D$-dimensional neutral fluid.

The EOMs for the neutral fluid consist of the conservation of energy-momentum tensor
\begin{equation}
\hat \nabla^N \hT^\fluid_{N M} (\hat x) = 0.
\end{equation}
The energy-momentum tensor is given by a derivative expansion of hydrodynamic fields, which in our case consists of parity-even terms up to the first order in gradients. It is given by two terms:
\begin{align}
\hat T_{M N}^{\fluid} (\hat x)
&=\hat T^\pf_{MN} (\hat x) +\hat T^{\text{diss}}_{MN} (\hat x)
\, ,
\label{eq:totalfluidT}
\end{align}
where $\hat T^\pf_{MN}$ is the perfect fluid part and $\hat T^\diss_{MN}$ contains the dissipative effects. In this work, we do not assume a priori an equation of state for the fluid, so we treat all hydrodynamic fields as independent. For later treatment, we find it convenient to use the vielbein formalism. The vielbein
$E_M{}^A$ is related to the metric as\footnote{Further developments on the incorporation of fermionic DOFs are allowed in this scheme, due to the vielbein formalism approach.}
\begin{equation}
\hat g_{MN}(\hat x) = E_M{}^A(\hat x) E_N{}^B(\hat x) \eta_{AB}
\,, \qquad \qquad \eta_{AB} = (-+\ldots+) \, .
\label{eq:metricvielbein}
\end{equation}
Thus,
\begin{align}
\hat T_{MN}^{\rm fluid}(\hat x)
&=
E_M{}^A(\hat x) E_N{}^B (\hat x)\left( \hat T_{A B}^\pf(\hat x) + \hat T_{A B}^\diss(\hat x) \right)
\, .
\label{fluidemtensor}
\end{align}

At zeroth-order in the gradient expansion, the fluid is perfect, so
\begin{align}
\hat T^\pf_{MN} (\hat x)
&=
[\heps (\hat x) + \hp (\hat x)] \hu_M (\hat x) \hu_N (\hat x) + \hp (\hat x) \hat g_{MN} (\hat x)
\, .
\end{align}

To study the dissipative part of energy-momentum tensor, it is necessary to specify the hydrodynamic frame. Dependence on the hydrodynamic frame arises as a consequence that the macroscopic variables that characterize the fluid do not have unique microscopic definitions. This permits us to have some freedom to select a convenient frame and therefore redefine them in a simple manner. A convenient choice to fix this arbitrariness uses the projection on the dissipative part of the energy-momentum tensor,
\begin{equation}
\hat u ^M \hat T^{\text{diss}}_{MN} =0 \, .
\end{equation}
This is known as the Landau frame. The dissipative part of energy-momentum that follows from this choice is
\begin{align}
\hat T^\diss_{AB} (\hat x)
&=
-2\hat\eta (\hat x) \, \hat \sigma_{AB} (\hat x) - \hat\zeta (\hat x) \, \hat \perp_{AB} (\hat x) \, \hat\theta (\hat x)
\, ,
\label{eq:Tdiss}
\end{align}
where $\hat \eta$ is the shear viscosity and $\hat \zeta$ is the bulk viscosity of the $D$-dimensional neutral fluid. We also denote the projection tensor along the velocity vector as $\hat \perp_{AB}$, the shear tensor as $\hat \sigma_{AB}$, and the expansion scalar as $\hat \theta $.
They are defined as follows:
\begin{align}
\hat\perp_{AB}(\hat x)
&= \hat\eta_{AB} + \hu_A (\hat x) \hu_B (\hat x)
\, ,
\nonumber\\
\hat \sigma_{AB} (\hat x)
&=
\hat \perp_{(A}{}^C (\hat x) \hat \perp_{B)}{}^D  (\hat x) \mathfrak{\hat D}_C \hu_D (\hat x)  - \frac{1}{\hp(x)}\hat \theta (\hat x) \, \hat \perp_{AB} (\hat x)
\, ,
\\
\hat \theta (\hat x) \, \, \, \,
&=
\mathfrak{\hat D}_A\hu^A (\hat x)
\, ,
\nonumber
\end{align}
where $\mathfrak{D}_A= E_A{}^M \partial_M + \hat \omega_A(\hat x) $ and $\hat\omega_A $ is the spin connection acting on the tangent frame.

 We minimally couple this $D$-dimensional neutral, dissipative fluid to the $D$-dimensional metric  $\hat g_{MN}$. The system is described by the $D$-dimensional Einstein equations sourced by the fluid,
\begin{align}
\hat R_{M N} (\hat x)
-\tfrac12 \hat g_{MN}(\hat x) \hat R (\hat x)
 \, = \, \hat T_{MN}^{\rm perfect} (\hat x)  + \hat T_{MN}^{\rm diss} (\hat x)
\, .
\label{eq:EinsteinHD}
\end{align}
It is important to stress that the energy-momentum tensor sourcing the Einstein's equation includes both ideal and dissipative parts. Nevertheless, while it is out of the scope of this work, we can apply the same approach to any order in the gradient expansion.

In the following section we are going to review the main aspects of the KK compactification of this system.

\subsection{Non-Abelian Kaluza-Klein reduction}
To construct non-Abelian hydrodynamics, we make use of the KK theory for dimensional reduction. In this section we sketch the main aspects of the KK compactification approach and the guidelines of our developments.

We start with the $D$-dimensional Einstein-fluid system given by Eq.~(\ref{eq:EinsteinHD}) and do a dimensional reduction on $n$-dimensional compact space $X_n$. We can effectively split the gravitational DOFs in $D$ dimensions into gravitational and additional DOFs in the $d=(D-n)$-dimensional reduced spacetime. The additional DOFs are scalar fields that characterize the size and shape of $X_n$ and, if the manifold admits Killing symmetries, vector fields with gauge symmetries. Likewise, we can split the fluid energy-momentum tensor in $D$ dimension into fluid's energy-momentum tensor and some vector currents in $d$-dimensional, reduced spacetime. Depending on the symmetry structure of Killing vectors on $X_n$, these vector currents can be either Abelian or non-Abelian. In this treatment, one must only keep a consistent truncation of light modes, setting the massive modes to zero. Consistency requires that heavy modes that are dropped are not sourced by the light modes one keeps.

As for gravity, it is known that KK dimensional reductions that involve Abelian isometries are always guaranteed to be consistent, as the heavy and light modes do not mix each other. It is also known that for some internal spaces (maximally symmetric spaces and group manifolds), reductions that involve non-Abelian isometries are consistent too. As for the matter,  KK compactification of a fluid without gravity (and hence, without dynamical gauge fields coupled to the fluid) on $n$-dimensional torus $X_n = \mathbb{T}^n$ was recently studied \cite{DiDato:2013cla}. The reduction led to a fluid carrying $U(1)^n$ ``global'' charges, and to relations between $D$-dimensional heat transport coefficients and $d$-dimensional, reduced charge transport coefficients. The results are in agreement with results known independently, so it suggests that KK reduction that involves Abelian isometries is consistent for the fluid as well.

The next step is the KK dimensional reduction of Einstein-fluid system on a group manifold $X_n = G$ \cite{Scherk:1979zr} of dimension $n = {\rm dim}(G)$. The group manifold $G$ is describable in terms of the Maurer-Cartan one-forms $\sigma^m$. These one-forms are invariant under left multiplications by a group element $g \in G$. Thus, this left multiplication is an isometry of the metric $g(G)$. The massless fields of the $d$-dimensional, system will be the metric $g_{\mu\nu}$ and the non-Abelian gauge fields with gauge group $G$. This is the main reason why we reduce the higher-dimensional Einstein-fluid system on group manifolds: the reduction naturally ``charges'' and couples the fluid to dynamical non-Abelian gauge fields. The reduction will translate the $D$-dimensional conservation laws into the $d$-dimensional, reduced conservation of both energy-momentum tensor and non-Abelian currents.

From the KK compactification, we obtain the system of colored fluid interacting with non-Abelian gauge theory. Nevertheless, the reduction also will bring in additional DOFs. Depending on the physical situations we are interested in, one may keep them as part of the system or truncate them out. For the formulation of non-Abelian hydrodynamics relevant for heavy-ion collision, we will only keep the non-Abelian gauge fields but none others. That is, we will decouple the gravitational DOFs by sending the Newton's constant to zero. This will put the non-Abelian hydrodynamics on $d$-dimensional Minkowski spacetime. We will also need to truncate the dilaton (that parametrizes the volume of $G$) and other scalar fields that emerge by setting them to be covariantly constant with respect to the non-Abelian gauge fields. Varying them, however, would result in change of the $d$-dimensional equations of state.

Let us stress that this approach does not rely on neither kinetic theory nor Lagrangian formulations. We are giving an {\sl ab initio} derivation of non-Abelian hydrodynamics by just assuming that a neutral fluid coupled to Einstein field equations is well-defined in $D$ dimensions.

Finally, let us comment on a technical caveat related to the Yang-Mills gauge group. In our approach, the KK dimensional reduction is done on the EOMs. This bears some consequences in the possible choices of the group manifold $X_n = G$. In particular, dimensional reduction of the EOMs allows for gauge groups whose structure constants are traceful, \emph{i.e.}, $f_{mn}{}^n\neq0$ (\emph{cf.} \cite{Roest:2004pk}).

\section{Charged Fluid Coupled To Maxwell Theory}
\label{sec:u1}

As a first step to introduce the technicalities that KK theory requires and build intuitions therein, we will consider the KK reduction of Einstein-fluid system on a group manifold with Abelian isometries. Thus, we choose the internal manifold to be a $n$-torus, $X_n=U(1)^n$. For simplicity, we will restrict ourselves to a perfect fluid and incorporate the dissipative effects in the next section.

Consider the KK reduction of a perfect fluid given by Eq.\eqref{eq:EinsteinHD} on a $\mathbb{S}^1$ internal manifold, where $\hat T^\fluid_{MN} = \hat T^\pf_{MN}$. We will show that the KK reduction gives rise to a charged perfect fluid interacting with Maxwell electromagnetism.
%

\subsection{Reduction on Abelian group manifold}
For the KK reduction on a circle, let us assume the following ansatz for the vielbein $E_M{}^A(\hat x)$ in Eq.~\eqref{eq:metricvielbein} as
\begin{align}
E_M{}^A(\hat x)
&=
\begin{pmatrix}
e^{\alpha\phi(x)} e_\mu{}^a(x) & e^{\beta \phi(x)} A_\mu(x)
\\
0 & e^{\beta \phi(x)}
\end{pmatrix}
\, .
\label{vielbein}
\end{align}
Curved indices of the $D$-dimensional spacetime will be split as $M=\{\mu, \, z\}$ whereas we will denote flat indices as $A=\{a, \, \underline{z}\}$. We will also assume that all the fields that appear in the ansatz only depend on the $d$-dimensional coordinates $x^\mu$ of ${\cal M}_d$~\footnote{In Section \ref{sec:su2}, we will assume some dependence on the internal coordinates, which will yield to non-Abelian gauge symmetry upon reduction.}. The dilaton $\phi(x)$, which measures the size of $X_n$, is weighed by the reduction-specific coefficients
\begin{align}
\alpha^2 =
\frac{n}{2(d+n-2)(d-2)}
\qquad \mbox{and} \qquad
\beta=
-\frac{(d-2)\alpha}{n}
\, .
\end{align}
Though in this section we evaluate $n=1$, we will keep $n$ generic.

Let us start by substituting the compactification ansatz into the $D=(d+1)$-dimensional Einstein field equations
\begin{align}
\hat G _{MN}(x)\equiv \hat R_{MN}(x) - \frac{1}{2}\hat g_{MN}(x) \hat R(x) = \hat T_{MN}^{\rm fluid}(x)
\, ,
\label{defining}
\end{align}
and recast the differential equations.
%
%
%
The components $\hat G_{\mu\nu}$, $\hat G_{\mu z}$, and $\hat G_{zz}$ give the $d$-dimensional gravitational, gauge, and dilaton field equations, respectively. Though we do not specify the structure of fluid energy-momentum tensor $\hat T_{MN}^{\rm fluid}$, we will return to it after analyzing the component equations.

The $\hat  G_{\mu z}$ components imply the Maxwell equations coupled to a current.
\begin{align}
& \nabla^\nu\left( 
Q_e^{-2}(x) F_{\mu\nu} \right) = Q^{-2}_e(x)
J^{\rm e}_\mu(x)  \, ,
\label{eq:EOMA}
\end{align}
where $Q_e$ is the dilaton-dependent gauge coupling,
\bea
Q_e(x) \equiv e^{(d-1)\alpha \, \phi(x)},
\eea
and the current is given by
\begin{align}
& J^{\rm e}_\mu (x) \equiv
2 e^{ (\beta + 2 \alpha 
)  \phi(x)} Q_e(x)  \, e_\mu{}^a(x) \hat T^{\rm fluid}_{a \uz}(x)
\, .
\label{eq:u1current}
\end{align}
Hence, the $\hat G_{\mu z}$ components of Einstein equations automatically define the electromagnetic dynamics of the system, including the current $J^{\rm e}_{\mu}(x)$ of the fluid. Thus, the fluid becomes charged whenever it has non-vanishing flow around $\mathbb{S}^1$. Being porportional to $T^{\rm fluid}_{a \uz}(x)$, the electric current $J^{\rm e}_\mu (x) $ will be proportional to the reduced velocity field $u_\mu(x)$. The dilaton field that measures the size of $\mathbb{S}^1$ has the effect of spacetime-dependent unit of electric charge, $Q_e(x)$. As discussed in the previous section, we take the KK reduction as an {\sl ab initio} approach for deriving consistent hydrodynamic equations. As such, we will eventually set the dilaton to be constant-valued.

This same pattern to the other components of \eqref{eq:EinsteinHD}. From the $\hat G_{\mu \nu}$ components, we obtain the $d$-dimensional Einstein equations sourced by the charged fluid, the $U(1)$ gauge field and the dilaton:
\begin{align}
G_{\mu \nu}(x) \equiv R_{\mu \nu}(x)
- \frac12 g_{\mu\nu}(x) R(x) = \Tt_{\mu\nu}(x)
\, ,
\label{einstein}
\end{align}
where the right-hand side defines the total energy-momentum tensor of the $d$-dimensional system
\begin{align}
& \Tt_{\mu\nu}(x)
=
T_{\mu\nu}^{\rm fluid}(x)
+ \left( \frac{1}{2}\partial_\mu\phi\partial_\nu\phi
-\frac{1}{4} g_{\mu \nu} (\partial\phi)^2 \right)(x)
+ \frac{1}{2} 
Q_e^{-2}(x) \left( F^2_{\mu \nu} - \frac{1}{4} F^2 g_{\mu\nu} \right)(x)
\, .
\end{align}
The last two terms are contributions of dilaton field and Maxwell field, while the first term is the energy-momentum tensor of charged fluid, defined by
\begin{align}
T_{\mu \nu}^{\rm fluid}(x) \equiv e^{2\alpha \phi(x)} e_\mu{}^a(x) e_\nu{}^b(x) \hat T_{a b}^{\rm fluid}(x)
\, .
\end{align}

Finally, let us consider the $\hat G_{zz}$ component. We obtain the $d$-dimensional dilation field equation, sourced by both the fluid and the Maxwell gauge field,
\begin{align}
\Box\phi (x) = 2 \alpha \, D(x)
\, ,
\end{align}
Again, the right-hand side of the equation defines the dilatation current,
\begin{align}
D (x) & \equiv
-  \frac{d-1}{4}\,  Q_e^{-2} (x) 
F_{\mu \nu}^2 (x)
+ (d-1)\, e^{2\alpha\phi (x)} \, \hat T_{\uz \uz}^{\rm fluid}(x) - e^{2\alpha\phi (x)} \,
{\hat T}^{\rm fluid} (x)
\, .
\label{eq:EOMphiU1}
\end{align}
where $\hat T^\fluid := \hat {T}^\fluid{}_M{}^M$ is the trace of the $D=(d+1)$-dimensional energy-momentum tensor.

The Einstein tensor in the defining equation Eq.~\eqref{defining} obeys the Bianchi identity, from which conservation laws of various currents we identified above are derived. The conservation laws on current $J^{\rm el}_\mu$ and total energy-momentum tensor $T^{\rm total}_{\mu \nu}$ result relevant for the Maxwell-plasma system. For the charge current, covariant divergence of Eq.\eqref{eq:EOMA} gives
\begin{align}
\nabla^\mu\left[\nabla^\nu\left( Q_e^{-2} 
F_{\mu\nu} \right)\right]
=
\nabla^{[\mu}\nabla^{\nu]} \left( Q_e^{-2}(x) 
\right)F_{\mu\nu}  = 0
\, ,
\end{align}
where we have used the torsion-free condition for $d$-dimensional spacetime. This implies
\begin{align}
\nabla^\mu\big(   e^{(\beta-(d-3)\alpha) \phi} e_\mu{}^a \hat T_{a \uz} \big)(x) = 0 \qquad
\rightarrow \qquad
\nabla^\mu(Q_e^{}(x) J^{\rm e}_\mu (x)) = 0
\, ,
\end{align}
which results the to the conservation law of electric current $ J^{\rm e}_\mu$, generalized by the dilaton field.

From the Bianchi identity of Einstein tensor in  Eq.~(\ref{einstein}) we obtain
\begin{align}
\nabla^\mu \Tt_{\mu\nu} (x) = 0
\, .
\end{align}
This implies that variations in the fluid energy-momentum tensor are balanced by the change of the Maxwell energy-momentum tensor and the dilaton.
\begin{align}
\nabla^\mu T^{\rm fluid}_{\mu \nu}(x)  = - \nabla^\mu \big( Q_e^{-2} [F^2_{\mu \nu} -{1 \over 4} F^2 g_{\mu \nu}] \big)(x).
\end{align}
On-shell, this conservation is equivalent to
\begin{align}
\nabla^{\mu} T^{\rm fluid}_{\mu\nu}
+ \big( e^{[\beta-(d-3)\alpha]\phi}  e_{\mu}{}^a \hat T^{\rm fluid}_{a \uz} \big) F_\nu{}^\mu  +
e^{2\alpha\phi}\big[
	(d-1)\alpha \hat T^{\rm fluid}_{\uz\uz} - \alpha \hat T^{\rm fluid}
	\big] \nabla_\nu \phi
  = 0
\, .
\label{eq:lorentzU1}
\end{align}
We interpret this as the generalization of the Lorentz force equation of Maxwell-plasma under the presence of the dilaton field. Once again, the role of the KK approach is just a tool to facilitate the {\sl ab initio} derivation of charged fluid interacting with Maxwell theory. Therefore, setting the dilaton to be constant-valued we obtain the standard form of the Lorentz force equation:
\begin{align}
\nabla^{\mu} T^{\rm fluid}_{\mu\nu}(x) = Q_e(x) {F^\mu}_\nu (x) J_\mu^{\rm el}(x)
\, .
\end{align}

\subsection{Abelian reduction of energy-momentum tensor}

So far, we have not made any assumption on the energy-momentum tensor $\hat T^{\fluid}_{MN}$ of the neutral fluid we started from. We now study $\hat T_{MN}^{\rm fluid}$ under a well-motivated ansatz for the higher-dimensional velocity field $\hat u(\hat x)$ and the other scalar quantities. To gain better intuition about physics, we will restrict the $D$-dimensional neutral fluid to a perfect fluid. In Section \ref{sec:su2}, we will consider the dissipative contributions.

The $D$-dimensional velocity field $\hat u^M$ has $(D-1)$ independent components, as it is conveniently normalized by Eq.~(\ref{eq:uu1HD}):
\begin{align}
\hu^M(\hat x) \hu^N(\hat x) \hat g_{M N}(\hat x) =  -1
\, .
\label{eq:UUcircle}
\end{align}
The ansatz that we will assume for the velocity field is:
\begin{align}
\begin{split}
\hu_a &=  u_a (x) \cosh\varphi (x) \, ,
\\
\hu_\uz &=  \sinh\varphi (x)
\, ,
\end{split}
\label{velocity}
\end{align}
where $u_a (x) $ is the velocity field of charged fluid in $d$ dimensions, which is normalized as $u^a(x) u^b(x) \eta_{ab} = -1$. The scalar field $\varphi (x)$ parametrizes the degree of freedom associated with the internal component of the velocity, $\hu_{\uz}$. 
%
%

Substituting the ansatze for the vielbein Eq.~(\ref{vielbein}) and the velocity fields Eq.~(\ref{velocity}) into the energy-momentum tensor, we will obtain the defining varibles of the $d$-dimensional fluid in terms of the $D$-dimensional ones. That is to say, we find that the energy-momentum tensor in $d$ dimensions is
\begin{align}
T_{\mu\nu}^\pf (x) &= e_\mu{}^a(x) e_\nu{}^b(x) \hat T_{ab}(x) \nonumber \\
&= (\epsilon+p)\, u_\mu(x) u_\nu(x) + p \, g_{\mu\nu}(x)
\, ,
\label{reduced-emtensor}
\end{align}
where the energy density $\epsilon(x)$ and the pressure $p(x)$ are given by
\begin{align}
\epsilon(x)
&=
e^{2\alpha\phi(x)}\left(
	\heps(x) \cosh^2\varphi (x)
	+ \hp (x) \sinh^2\varphi (x)
	\right)
\, , &
p (x) &= e^{2\alpha\phi (x)} \hp (x)
\, .
\end{align}

By substituting the velocity ansatz Eq.~\eqref{velocity} into Eq.~\eqref{eq:u1current}  
we obtain the electric current
\color{black}
\begin{align}
J^{\rm e}_\mu(x)
\equiv&
\,  2\, (\epsilon (x)+p (x)) \, e^{[\beta +(d+1)\alpha]\phi(x)} \, \tanh \varphi (x)\, u_\mu(x)
\, .
\end{align}
As anticipated, the charge current is proportional to the velocity field $u_\mu$.
Again, let us analyze the case for which the dilaton field is constant. Then, the energy-momentum conservation, Eq~\eqref{reduced-emtensor} 
leads to
\begin{align}
\nabla^\mu T^\pf_{\mu \nu}(x) = Q_e(x) F^\mu{}_\nu (x) J_\mu^{\rm e}(x)
\, .
\end{align}
This is precisely the Lorentz force equation we have directly derived from the reduction of the Einstein-fluid system in the last section.
%

One can straightforwardly generalize the above construction by taking the internal space $X_n$ to be an $n$-torus $\mathbb{T}^n$. It will give rise to a fluid charged under $n$ independent Abelian electromagnetic fields with $U(1)^n$ gauge symmetry.

After analyzing the system of a fluid charged under Abelian gauge fields, we will address the case for which the gauge symmetry is non-Abelian. To carry out this problem, the internal manifold will be a group manifold whose isometry group is non-Abelian. We will choose SU(2) for simplicity but the procedure applies to any other gauge group.

\section{Colored Fluid Coupled To Yang-Mills Theory}
\label{sec:su2}

We now construct non-Abelian hydrodynamics of Yang-Mills plasma. Here, our goal is to derive \emph{ab initio} the EOMs of a dissipative fluid carrying non-Abelian SU(2) charges and interacting with Yang-Mills theory. To do so, our idea is again to start with an Einstein-fluid system in $D$ dimensions Eq.\eqref{eq:EinsteinHD} and perform a KK dimensional reduction on a SU(2) group manifold \cite{Scherk:1979zr} (for a review, \emph{cf.} \cite{Pope:web,Roest:2004pk,Dibitetto:2012xd}). After the reduction, we will find an SU(2) colored fluid interacting with SU(2) Yang-Mills theory in $d$ dimensions. As SU(2) group manifold is three-dimensional, our setup corresponds to $n=3$ and hence $D= d+3$.
Nevertheless, this method can be applied to any group manifold $G$, having thus a colored fluid interacting with Yang-Mills theory of gauge group $G$.

\subsection{Compactification on SU(2) group manifold}
Let us consider the following KK ansatz for the $D$-dimensional vielbein:
\begin{align}
E_M{}^A
&=
\begin{pmatrix}
e^{\alpha\phi} \hat e_{\mu}{}^a	&  e^{\beta\phi} \hat A_\mu{}^p \hat e_{p}{}^{ \alpha}
\\
0									& g^{-1} e^{\beta\phi} \hat e_m{}^\alpha
\end{pmatrix}
\, ,
\label{eq:ansatzsu2}
\end{align}
where
\begin{align*}
\hat e_\mu{}^a(\hat x) &= e_\mu{}^a(x)
\, ,
\\
\hat e_m{}^\alpha(\hat x) &= \fu_m{}^{n}(y) \cV_n{}^\alpha(x)
\, ,
\\
\hat A_\mu{}^m(\hat x) &= (\fu^{-1})_n{}^m(y) A_\mu{}^n(x)
\, .
\label{components}
\end{align*}
and $g$ is a coupling constant $g$. As in the Abelian reduction, we split the curved manifold indices as  $M = (\mu, m)$ where $\mu=1\ldots d$ and $m=1\ldots n$, tangent space indices as $A=\{a, \, \alpha\}$, where $a=1\ldots d$ and $\alpha=1\ldots n$, and local coordinates as $\hat x^M = (x^\mu, y^m)$. For the SU(2) case, $n=3$. These group-valued indices can be freely lowered and raised without loss of generality.

In various Weyl factors, the dilaton field $\phi(x)$ is weighed with the coefficients
\begin{align}
\alpha^2
=
\frac{n}{2(d+n-2)(d-2)}
\qquad \mbox{and} \qquad
\beta
=
-\frac{(d-2)\alpha}{n}
\, .
\end{align}

The matrix $\fu_m{}^n(y)$ in Eq.(\ref{components}) is a twist field that carries the information of the SU(2) group manifold. After the reduction, this information is encoded in the $d$-dimensional system through the structure constants,
\begin{align}
f_{mn}{}^p := -2 (\fu^{-1})_m{}^s(y) (\fu^{-1})_n{}^t(y) \partial_{[s} \fu_{t]}{}^p(y)
\, .
\end{align}
Though the twist matrix field ${\fu_m}^n(y)$ varies over the group manifold (hence depends on the internal coordinates $y$), the combination on the r.h.s. of this equation needs to be constant-valued in order for them to be the structure constants of the Lie algebra associated with the group manifold.

The ansatz can be explicitly expressed in terms of the Maurer-Cartan one-forms $\sigma^m$ of the SU(2) group  manifold by combining the fields as
\begin{align}
\hat E^a(x)
&=
e^{\alpha\phi (x)} e^a(x)
\, ,
\nonumber \\
\hat E^\alpha (x)
&=
g^{-1} e^{\beta\phi (x)} \cV_m{}^\alpha(x) (\boldsymbol \sigma^m - g \mathbf A^m(x))
\, ,
\end{align}
where $\boldsymbol \sigma^m \equiv u_n{}^m \rmd x^n$ is the left-invariant one-form of $G$, satisfying the Maurer-Cartan equation
\begin{align}
\rmd \boldsymbol \sigma^m + \frac12 f_{np}{}^m \boldsymbol \sigma^n\wedge \boldsymbol \sigma^p = 0
\, ,
\end{align}
and thus $f_{np}{}^m$ are the structure constants of the isometry group $G$ of the internal manifold.

Before carrying out the non-Abelian reduction on the group manifold $G$, we introduce new notations for the physical variables in $d$ dimensions. We shall build from the scalar vielbein $\cV$ two scalar metrics
\bea
{\bf M}_{m n}=\cV_m{}^\alpha \cV_n{}^\beta \delta_{\alpha\beta} \qquad \mbox{and}
\qquad
\mathbb{M}^{\alpha\beta} \equiv \cV_m{}^\alpha \cV_n{}^\beta \delta^{mn},
\eea
which are SU(2) invariant and SU(2) covariant, respectively.
We denote the trace as $\mathbb M \equiv \mathbb M_{\alpha}{}^\alpha$. We define the covariant derivatives $D_\mu({\bf A})$ and $\cD_\mu(\cV)$ as
\begin{align}
\begin{split}
D_\mu \cV_{m}{}^\alpha
& \equiv
\partial_\mu \cV_{m}{}^\alpha - g {\bf A}_{\mu \, m}{}^n 
\cV_{n}{}^\alpha
\, ,
\\
\cD_\mu \cV_m{}^\alpha
& \equiv
D_\mu \cV_m{}^\alpha + \, \, {\mathbb{Q}_{\mu}}^\alpha{}_\beta \cV_m{}^\beta
\, ,
\end{split}
\end{align}
where the elementary gauge field used in $D_\mu$ is given by
\bea
{\bf A}_{\mu \, m}{}^n (x) \equiv A^p{}_{\mu} (x) f_{pm}{}^n
\eea
and the composite gauge fields used in ${\cal D}_\mu$ are built from the scalar vielbein as
\begin{align}
\begin{split}
\mathbb P_{a\, \alpha\beta}(x)
&\equiv
e_a{}^\mu \mathbb P_{\mu\, \alpha\beta}
=
\frac12\left(
	\cV_\alpha{}^m\, D_a({\bf A}) \cV_m{}^\beta
	+\cV_\beta{}^m\, D_a({\bf A}) \cV_m{}^\alpha
	\right)
\, , \\
\mathbb Q_{a\, \alpha\beta}(x)
&\equiv
e_a{}^\mu \mathbb Q_{\mu\, \alpha\beta}
=
\frac12\left(
	\cV_\alpha{}^m\, D_a({\bf A}) \cV_m{}^\beta
	-\cV_\beta{}^m\, D_a({\bf A})  \cV_m{}^\alpha
	\right)
\, .
\end{split}
\end{align}
The distinction is that, while $D_\mu$ is the ordinary gauge covariant derivative, $\cD_\mu$ accounts for quantities that are adjoined by the scalar vielbein $\cV_m{}^\alpha$. Finally, the Yang-Mills field strength two-form ${\bf F}^m$ of ${\bf A}^m$ is defined as
\begin{align}
{\bf F}^m
&\equiv
\rmd {\bf A}^m + \frac 12 \, g \, f_{n p}{}^m \,  {\bf A}^n\wedge  {\bf A}^p
\, .
\end{align}
This field strength typically appears dressed up by the scalar fields, so we also denote the tangent space (both in internal and spacetime manifolds) field strength two-form as ${\bf F}^\alpha{}_{ab} \equiv \cV_m{}^\alpha {\bf F}^m{}_{ab}$.

\subsection{Field equations for Yang-Mills plasma}

To obtain the EOMs of the $d$-dimensional system we will substtute the ansatz Eq.~\eqref{eq:ansatzsu2} into Einstein equations and recast the resulting expressions~\footnote{Details of the calculations for extracting the equations of motion are relegated to Appendix \ref{app:su2details}.}.

Let us start with the EOMs for the SU(2) gauge fields. They descend from the $\hat G_{\mu n}$ components in Eq.\eqref{eq:EinsteinHD}. Working in the tangent space we obtain
\begin{align}
\cD^b(
Q_c^{-2}(x) {\bf F}^\beta{}_{ab}(x)) + 
Q_c^{-2}(x) \mathbb P^{b \beta}{}_{\gamma}(x)
\, {\bf F}^{\gamma}{}_{a b}(x)
=  
Q_d^{-2} (x) {\bf J}^\beta_a (x)
\, ,
\label{eq:EOM-gauge}
\end{align}
where
\bea
Q_c(x) := e^{\frac13\alpha(d+1)\phi(x)}
\eea
is the dilaton-dependent gauge coupling, and
\begin{dmath}
{\bf J}_{a \beta}(x) = 2 
\left[g\, Q_c^2 (x) \, \epsilon_{\beta\gamma\delta}  {\mathbb P_a}^\gamma{}_\lambda(x) \, \mathbb M^{\delta\lambda}(x)
-   \, Q_c^{}(x) \left(e^{2 \alpha \phi(x)} \, 
\hat T^{\rm fluid}_{a \beta}(x) \right) \right]
\, .
\label{colorcurrent}
\end{dmath}
is the color current.
For covariantly constant scalars, Eq.\eqref{eq:EOM-gauge} is reduced to
\begin{align}
\cD^b( {\bf F}^\beta{}_{ab})(x) = \mathbf J^\beta_a(x) \, ,
\end{align}
which is the standard form of the Yang-Mills field equations coupled to color current.

The Einstein field equations descend from the $\hat G_{\mu\nu}$ components:
\begin{align}
G_{\mu \nu}(x) = R_{\mu\nu}(x) - \frac{1}{2}g_{\mu\nu}(x) R(x) = T^{\tot}_{\mu\nu}(x)
\, ,
\label{eq:EOMeins}
\end{align}
where $T^{\tot}_{\mu\nu} := e_\mu{}^a e_\nu{}^b T^{\tot}_{ab}$ is the total energy-momentum tensor, with
\begin{align}
T^{\tot}_{ab}(x)
&= e^{2\alpha\phi(x)} \hat T^{\rm fluid}_{ab}(x) + \tfrac{1}{2} Q_c^{-2}(x) 
\left(
	\eta^{cd} {\bf F}^\alpha{}_{a c} (x) {\bf F}^\alpha{}_{b d}(x)
	- \tfrac12\eta_{ab} ({\bf F}^\gamma)^2(x) \right)
\nonumber \\
& + \tfrac12 \left(\partial_{a}\phi(x) \partial_b\phi(x) - \tfrac12 \eta_{ab} (\partial\phi)^2(x) \right)
+ (\mathbb P_{a\beta\gamma}(x) \mathbb P_{b \beta\gamma}(x) - \tfrac12 \mathbb P^2(x) \eta_{ab})
\nonumber \\
& -g^2 \, Q_c^2(x) 
\left(
	\mathbb M^{\gamma\delta}(x) \mathbb M^{\gamma\delta}(x)
	- \tfrac12 \mathbb M^2(x)
	\right)\eta_{ab}
\, .
\label{eq:total-et}
\end{align}
From the first line, we read off the energy-momentum tensor $T^{\rm fluid}_{ab}$ of the colored fluid:
\begin{align}
T^{\rm fluid}_{ab}(x) \, =  \, e^{2\alpha\phi(x)} \hat T^{\rm fluid}_{ab}(x)
\,.
\label{eq:effectiveTsu2}
\end{align}

Other field equations also yield relevant information on currents and their conservation laws. The equation of motion for dilaton field is obtained from the trace of Eq.\eqref{eq:EinsteinHD},  $\hat G_m{}^m$:
\begin{align}
\Box \phi(x)
\, = \, {1 \over 2 \alpha} D(x)
\, ,
\end{align}
where $D(x)$ is the dilation current 
\begin{dmath}
D(x) = - \frac{1}{2(d-2)} Q_c^{-2} (x) 
({\bf F}^{\alpha})^2(x)
+ \frac{2g^2}{d-2} Q_c^2(x) 
(\mathbb M^{\alpha\beta}(x) \mathbb M^{\alpha\beta}(x) - \tfrac12 \mathbb M^2(x))
- \frac{2}{d-2} e^{2\alpha\phi(x)} \Big(
	\frac{3}{d+1} \hat T^{\rm fluid}(x) - \hat T^{\rm fluid}_{\alpha\beta}(x) \delta^{\alpha\beta}
	\Big)\, .
\label{eq:EOMphiSU2}
\end{dmath}
The first line is the contribution of SU(2) gauge fields and scalar fields, whereas the second one is the contribution of colored fluid. As we can check, there is no non-linear contribution of the dilaton field itself apart from the Weyl factors.

The equation of motion for the group-valued scalar fields $\cV_m{}^\alpha(x)$ is given by a linear combination of the $\hat G_{mn}$ components  and the trace $\hat G_m{}^m$:
\begin{align}
\cD^a (\mathbb Q) \mathbb P_{a \alpha\beta}
= \mathbb J_{\alpha \beta}
\, ,
\end{align}
where
\begin{dmath}
\mathbb J^{\alpha \beta}(x) = \tfrac12 Q_c^{-2}(x) 
\left[
	{\bf F}^\alpha{}_{ab}(x) {\bf F}^{\beta}{}_{cd}(x) \eta^{ac} \eta^{bd}
	- \tfrac13 ({\bf F}^\gamma)^2 \delta^{\alpha\beta}
	\right]
	+ 2 e^{2\alpha\phi}\left[
	\frac13 
	\hat T^{\rm fluid}
	\delta_{\alpha\beta}
 	- \hat T^{\rm fluid}_{\alpha\beta}
	\right]
	\\
	\\
+4g^2 \, Q_c^2(x) 
\left[
	\mathbb M^{\alpha\gamma} \mathbb M^{\beta\gamma}
	-\tfrac12 \mathbb M^{\alpha\beta} \mathbb M
	-\tfrac13 \left(
		\mathbb M^{\gamma\delta} \mathbb M^{\gamma \delta}
		-\tfrac12 \mathbb M^2
		\right) \delta_{\alpha\beta}
	\right] \, .
\label{eq:EOMVSU2}
\end{dmath}
The first line of this expression is the contribution of SU(2) gauge fields and colored fluid, while the last line corresponds to the contribution of group-valued scalar fields.


\subsection{Conservation laws}

The non-Abelian reduction of the Einstein-fluid system has led to a Yang-Mills plasma, consisting of colored fluid interacting with non-Abelian gauge fields (and also coupled to gravity, dilaton and group-valued scalar fields). 
In this section, we will further investigate the conservation laws of the system.

Likewise in Section \ref{sec:u1} for Maxwell plasma, we have not made any assumption on gravity and scalar fields so far. Nevertheless, in order to study the conservation of the simplest model for Yang-Mills plasma, we will truncate the system so that the $d$-dimensional metric is flat and scalar fields are covariantly constant. Such truncations will impose some constraints on the corresponding field equations of $\phi$ and $\cV_m{}^\alpha$, namely, Eqs.~\eqref{eq:EOMphiSU2} and \eqref{eq:EOMVSU2}. For this truncation to be consistent, we would need to solve these constraints. They will in turn impose some conditions on the $d$-dimensional Einstein equations\footnote{As for gravity, we can decouple the DOFs associated to the metric by taking the limit $G_N\rightarrow0$ for the Einstein equations.}  and the Yang-Mills field equations through Weyl factors and scalar potentials. In this section, we will simply consider the simplest consistent solution of these scalar fields, but will not explore the arena of possible non-trivial solutions. Nevertheless, it should be interesting to look into the implications of such nontrivial solutions (and their stability) in the context of fluid/gravity duality. It will also be important to understand to what extent these solutions constrain the values of the transport coefficients and other quantities that characterize the lower-dimensional fluid.


Firstly, let us analyze the color currents of the system and their conservation laws. The SU(2) Yang-Mills field equation Eq.\eqref{eq:EOM-gauge} can be recast:
\begin{align}
	D^b\big( Q_c^{-2}(x) \,
		\mathbf{M}_{nm}(x) {\bf F}^m{}_{ab}(x)
		\big)
	-2g \,  \epsilon_{\gamma\delta \sigma} {{\mathbb P}_a}^\gamma{}_\lambda {\mathbb M}^{\delta\lambda}  \cV_n{}^\sigma
	+ 2 Q_c^{-1}(x) \, (e^{2 \alpha \phi(x)} \, 
	\hat T^{\rm fluid}_{a \beta}) \cV_n{}^\beta
= 0
\, .
\end{align}
Applying a covariant derivative $D_a$, we obtain
\begin{align}
D^a\left(
	-2g \, \epsilon_{\gamma\delta\sigma} {{\mathbb P}_a}^\gamma{}_\lambda {\mathbb M}^{\delta\lambda}  \cV_n{}^\sigma
	+ 2 Q_c^{-1}(x) \left(e^{2 \alpha \phi(x)} 
	\, \hat T^{\rm fluid}_{a \beta}(x) \right) \cV_n{}^\beta (x)
\right)
= 0
\, .
\label{eq:currentconserv}
\end{align}
This reproduces the covariantly conserved color current Eq.(\ref{colorcurrent})
\begin{align}
{\bf J}_{m \, a}^{\rm color}
& =
	\left[2g \, Q_c^2(x) \, \epsilon_{\gamma\delta\sigma} {{\mathbb P}_a}^\gamma{}_\lambda (x) {\mathbb M}^{\delta\lambda}(x) 
	- 2  Q_c^{}(x) \left(e^{2 \alpha \phi(x)} 
	\hat T^{\rm fluid}_{a \sigma}(x) \right) \right] \cV_m{}^\sigma
	\, .
\label{eq:su2current}
\end{align}
The interpretation is clear: the first term is the color current sourced by the group-valued scalar fields, while the second term is the color current sourced by the colored fluid itself.
Being the non-Abelian counterpart of the $U(1)$ charged current,  the second term is proportional to the off-diagonal block of the energy-momentum tensor, $\hat T^\fluid_{a\beta}$. This block is non-zero if the $D$-dimensional fluid flows on the group manifold, so ${\bf J}^{\rm color}_{ma}$ is proportional to the internal velocity fields $u_a$.

Secondly, let us analyze the heat current of the Yang-Mills plasma and their conservation laws.
We already discussed that the Bianchi identity $\nabla^\mu G_{\mu\nu}=0$ of the $d$-dimensional Einstein equation, Eq.\eqref{eq:EOMeins} leads to the conservation of the total energy-momentum tensor
\begin{align}
\nabla^\mu T^\tot_{\mu\nu}=0
\, .
\end{align}
We would like to obtain the relations that this condition imposes among the $d$-dimensional degrees of freedom. Applying a covariant divergence on the total energy-momentum tensor Eq.\eqref{eq:total-et} and substituting the field equations of the Yang-Mills fields and scalar fields, we are left with an expression that involves first derivatives of the scalar fields and components of the energy-momentum tensor $\hat T^\fluid_{MN}$\footnote{We relegate details of the calculation to Appendix \ref{app:su2details}.}. This expression is the non-Abelian generalization of the Lorentz force, which involves not only the Yang-Mills field strength but also the group-valued scalar fields. Nevertheless, if we set these scalar fields to be covariantly constant, $D_a \cV_m{}^\alpha = D_a \varphi = 0$, we obtain
\begin{align}
D^a T_{ab}^{\rm fluid}
+2 Q_c^{-1}(x) e^{2 \alpha \phi(x)} 
\hat T_{c\alpha} \cV_n{}^\alpha {\bf F}^n{}_{bc}
=
e^{2\alpha\phi}\left(
	D^a \hat T_{ab}^{\rm fluid}   +
	2 Q_c^{-1}(x) 
	\hat T_{c\alpha} {\bf F}^\alpha{}_{bc}
	\right)
=0
\, .
\label{eq:lorentzsu2}
\end{align}
\emph{I.e.}, we get the standard expression of Lorentz force for Yang-Mills plasma:
\bea
D^a T_{ab}^{\rm fluid}(x)   = Q_c(x)  {\bf F}^{\alpha \, a}{}_b(x) {\bf J}^{\rm color}_{\alpha\, a}(x).
\eea

After doing the KK reduction of gravity sourced by a generic fluid $\hat T^\fluid_{MN}$, we are going to evaluate $\hat T^\fluid_{MN}=(\hat T^\pf+\hat T^\diss)_{MN}$ and study in detail the resulting $d$-dimensional fluid.

\section{Colored Fluid from Non-Abelian Reduction}

In this section, we will implement the KK compactification of the fluid energy-momentum tensor to construct the colored fluid and read off its defining variables.

\subsection{Non-Abelian Reduction of Fluid }
The energy-momentum tensor and the defining variable of the $d$-dimensional fluid will be read off after inserting the compactification ansatze for the veilbein and the rest of the expressions into the EOMs of the $D$-dimensional system.

For the non-Abelian reduction of the velocity fields $\hu^A$, we will assume an ansatz such that none of its components depend on the coordinates of the internal group manifold $G$. We can parametrize them as follows 
\begin{align}
\begin{split}
\hu^a &=   u^a(x) \, \cosh \varphi(x)
\, ,
\\
\hu^\alpha &=  {\bf n}^\alpha (x) \sinh\varphi  (x)
\, ,
\end{split}
\end{align}
where
\bea
u^a u^b \eta_{ab} = -1 \qquad \mbox{and} \qquad {\bf n}^\alpha {\bf n}^\beta \delta_{\alpha \beta} = 1.
\eea
The $d$-dimensional velocity has $(d-1)$ independent components, and the $n$-dimensional unit vector ${\bf n}$ has $(n-1)$ independent components. In total, along with $\varphi$, there are $(d-1) + (n-1) + 1 = D-1$ independent components.
The angular variable $\varphi$ measures the relative magnitude between the external and ``internal'' velocity fields. The unit vector $u^a$ is the boost in external sapcetime, while the unit vector $\mathbf n$ is the boost in the internal group manifold. They all fluctuate in external spacetime.

With this ansatz, we will now study the $d$-dimensional energy-momentum tensor of the fluid, Eq~\eqref{eq:totalfluidT}.

%
%
%
%
%
%
%
%

\subsection{Perfect colored fluid}
\label{sec:su2pf}

Firstly, we are going to characterize the colored perfect fluid in $d$ dimensions. This will allow us to identify its thermodynamic and scalar quantities in terms of quantities in $D$ dimensions.

The energy-momentum tensor of the $d$-dimensional perfect colored fluid is given by
\begin{align}
T^\pf_{ab}(x) = \left[\epsilon(x)+p(x)\right]
u_a(x) 
u_b(x) + p(x) \, \eta_{ab}
\, ,
\end{align}
where, using Eq.~\eqref{eq:effectiveTsu2}, the quantities are related to the $D$-dimensional ones as
\begin{align}
\begin{array}{rcl}
p(x)
&=&
e^{2\alpha\phi (x)} \hp(x)
\, ,
\\
\epsilon (x)
&=&
e^{2\alpha\phi (x)}\left[\,
	\cosh^2 \varphi (x) 
	\heps(x) + \sinh^2 \varphi (x) 
	\hp (x) \,
	\right]
\, .
\end{array}
\end{align}
From this, we find the speed of sound, $c_s$, in the perfect colored fluid as
\begin{align}
c_s^2
\equiv
\frac{\partial p}{\partial\epsilon}
=
\frac{1}{\cosh^2 \varphi (x) 
(\hat c_s^{-2}-1)+1}
\, , \qquad \mbox{where} \qquad
\hat c_s^{\, 2}= \frac{\partial \hp}{\partial \heps} \, .
\end{align}
The faster the fluid is boosted inside the group manifold, the slower the sound speed of the colored fluid.

The boost inside the group manifold generates the color current. From the current ${\bf J}^{\rm color}_{ma}$,  Eq.~\eqref{eq:su2current}, we 
have
\begin{align}
{\bf J}^{\rm color}_{ma}(x) = 
Q_c^{}(x) 
 {\mathfrak Q}_m (x) \, u_a (x)
\, .
\end{align}
Here, $\mathfrak Q_m$(x) is the color charge density attached to the fluid, which is defined as
%
\begin{align}
\mathfrak Q_m (x)
&=
2
{(\epsilon(x) +p(x) )}
\cV_m{}^\alpha(x) {\bf n}_\alpha(x) \tanh \varphi (x) 
\, .
\end{align}
%

\subsection{Entropy current}
The $D$-dimensional neutral fluid has entropy density $\hat s$, so the entropy current is given by
\bea
\mathfrak{\hat J}^{\hat s}{}_A = \hat s \hat u_A,
\eea
In the perfect fluid limit, the entropy current is covariantly conserved
\begin{align}
\hat \nabla^{M} \mathfrak{\hat J}^{\hat s}{}_M = 0
\, .
\label{entropy conservation}
\end{align}
From the ansatz Eq.~\eqref{velocity}, the entropy in $d$ dimensions is given by
\bea
s  = e^{2 \alpha \phi} \, \hat s \, \cosh \varphi
\, ,
\eea
and the entropy current in $d$ dimensions is given by
\bea
\mathfrak{J}_\alpha^s = s (x) \, {\bf n}_\alpha (x) \tanh \varphi (x),    \qquad
\mathfrak{J}_\mu^s(x) = s(x) \, u_\mu (x).
\eea
The conservation law Eq.(\ref{entropy conservation}) is reduced to
\begin{align}
\nabla^\mu \mathfrak{ J}^{s}{}_\mu = 0
\, .
\end{align}
where we have used the spin connection components of Appendix \ref{app:su2details}.

The neutral perfect fluid in $D$ dimensions satisfies the thermodynamic relation
\begin{align}
\heps + \hp = \hat \cT \hat s
\, ,
\end{align}
where $\hat \cT$ is the temperature. After the reduction, the $d$-dimensional fluid is colored, so its thermodynamic relation must account for the chemical potentials ${\boldsymbol \mu}^{\rm color}_m$ associated to the charges $\mathfrak Q_m$ in the form
\begin{align}
\epsilon+ p = \cT s + \mathfrak Q^m {\boldsymbol \mu}^{\rm color}_m
\, .
\end{align}
Requiring this Euler relation to hold in $d$ dimensions, we obtain that the $d$-dimensional temperature and chemical potentials are given by
\begin{align}
\cT (x) &= 
\hat \cT (x) \, {1 \over \cosh \varphi (x)} \,
\, ,
\nonumber
\\
{\boldsymbol \mu}^{\rm color}_m (x)
&=
 {\bf n}_\alpha (x) \cV^\alpha{}_m (x) \tanh \varphi (x)
\, .
\end{align}

So far, we have described the $d$-dimensional perfect fluid carrying non-Abelian SU(2) charges and given all its defining quantities in terms of the $D$-dimensional neutral fluid parameters. These results are in full agreement with the ones obtained for the Abelian case in Section 4. Built upon these consistency checks, we are going to consider dissipative effects of the fluid in the next section.

\subsection{Non-Abelian dissipative fluid}
\label{sec:su2diss}

We are going to extend our previous analysis by considering the dissipative part of energy-momentum tensor, $\hat T^\diss_{MN}$. This piece is given by
\begin{align}
\hat T^\diss_{AB}
&=
-2\hat\eta \hat \sigma_{AB} - \hat\zeta \hat \perp_{AB} \hat\theta
\, .
\end{align}
The correction of first-order in derivatives in $\hat T^\diss_{AB}$ will generate terms of first-order derivatives of the components of velocity fields $\hu_A$. Being velocity fields, these terms play the same role as second-order derivative of ordinary fields. Therefore, we will eliminate the derivatives by using their equations of motions, namely, the conservation laws.

In particular, if we consider Eqs.~\eqref{eq:currentconserv} and \eqref{eq:lorentzsu2}, we obtain
\begin{align}
u^\mu(x) \hat \nabla_\mu \varphi(x)  
&=
c_s^2 (x) \, \theta (x) \tanh \varphi(x)  
\, ,
\end{align}
where $\theta (x) \equiv \nabla_\mu u^\mu (x)$. Moreover, $\theta(x)$ is related to $\hat\theta\equiv \hat \nabla_M \hat u^M(x)$ by
\begin{align}
\hat \theta (x) = 
\cosh \varphi(x) \left(
	\theta 
	+ {\bf n}^\alpha  u^\mu 
	\hat\nabla_\mu {\bf n}_\alpha 
	\right)(x)
\, ,
\end{align}
so that when substituting, we have
\begin{align}
\hat \theta (x) = \cosh^3 \varphi(x)  \left( {c_s^2(x) \over \hat c_s^2 (x)} \right) 
\theta(x) 
\, .
\end{align}

In addition, the $d$-dimensional acceleration $a_\mu\equiv u^\nu \hat \nabla _\nu u_\mu$ is given by
\begin{align}
a_\mu
&=
\frac{{\rm sech}^2 \varphi (x)}{e^{2\alpha\phi(x)}\heps(x)+p(x) }
\hat \nabla_\mu \left( {1 \over p(x)} \right) 
+ n\theta(x) c_s^2(x) u_\mu(x)
\, ,
\end{align}
where
\begin{align}
\hat \nabla_\mu \left( {1 \over p(x)} \right)
&=
\frac{ e^{2\alpha\phi(x)}\heps(x) + p(x)}{2p^2(x)}  
\sinh 2\varphi(x) 
\hat\nabla_\mu \varphi (x)
\, .
\end{align}

With these results, we can estimate the $d$-dimensional coefficients associated with the dissipative terms.  For the $D$-dimensional neutral fluid, the shear and bulk viscosities can be read off from $\hat T^\diss_{AB}$. This occurs due to the fact that the fluid is described in the Landau frame, \emph{i.e.},
\begin{align}
\hat u^A \hat T^\diss_{AB}
= 0
\, .
\end{align}
Upon the non-Abelian KK dimensional reduction, the rearrangement of DOFs into $d$-dimensional Lorentz covariant representations implies that the reduced ones do not satisfy the Landau frame condition. In particular, we obtain
\begin{align}
u^a(x) \hat T^\diss_{ab}(x) + {1 \over \cosh \varphi(x)} \,  \hat u^\alpha (x) \hat T^\diss_{\alpha b} (x) = 0
\, ,
\end{align}
which straightforwardly leads to $u^a T^\diss_{ab} \neq 0 $.

On account of the frame-dependent structure of the energy-momentum tensor, departure from the Landau frame means that we cannot read off the $d$-dimensional transport coefficients associated with the dissipative terms from $T^\diss_{\mu\nu}$. To correctly identify these coefficients, we need a frame-invariant formulation of the dissipative terms. In addition, according to the second law of thermodynamics, it has to be guaranteed that the entropy current $\mathfrak{J}^s{}_a$ satisfies $\nabla^\mu \mathfrak{J}^s{}_\mu \ge 0$. Such frame-invariant description was developed in \cite{Bhattacharya:2011tra} for a fluid charged under an Abelian gauge field $A_\mu$. Here, we propose a generalization.

Using the frame-invariant approach as a guiding principle and also based on the gauge covariance of SU(2) group-valued quantities, we formulate the following expressions for the transport coefficients in the presence of non-Abelian gauge fields ${\bf A}^{m}{}_\mu$:
\begin{align}
\begin{split}
 \left(
	\perp_a{}^c(x) \perp_b{}^d(x)
	- \frac{1}{d-n} \perp_{ab}(x) \perp^{cd}(x)
	\right) T^\diss_{cd}(x)
&= -2 \eta (x) \, \sigma_{ab}(x)
\, ,
\\
 \frac{\partial p(x)}{\partial \mathfrak Q_m} u^a(x) 
{\bf J}^\diss{}_{a m}(x)
+ \left(
	\frac{1}{d-n}\perp^{ab}
	-\frac{\partial p}{\partial \epsilon} u^a u^b
	\right)(x) T^\diss_{ab}(x)
&=
- \zeta (x) \theta (x)
\, ,
\\
\perp_a{}^{b}(x)\left(
	{\bf J}^\diss{}_{b m}
	+ \frac{\mathfrak Q_m}{\epsilon+p} u^c T^\diss_{c b}
	\right)
-
{\boldsymbol \kappa}_{mn}(x) \left(
	-\perp_a{}^b D_b\left( \frac{{\boldsymbol \mu}_n}{\cT}\right)
	+ \frac{1}{\cT}{\bf F}^n{}_{ab} u^b
	\right)
&=
0
\, ,
\end{split}
\end{align}
where ${\bf J}^\diss{}_{a m} $ follows from  Eq.~\eqref{eq:su2current} using $\hat T_{MN}= \hat T^\diss_{MN}$,
$\boldsymbol{\kappa}_{mn}$ is the non-Abelian conductivity tensor, and $\eta$, $\zeta$, $\sigma$ are the $d$-dimensional dissipative coefficients.

At this stage, in order to obtain the effective dissipative coefficients, we need to substitute the expressions that we obtained for ${\bf J}^\diss_{am}$ and $\hat T^\diss_{a b}$ and work out these three equations.\footnote{In Ref. \cite{DiDato:2013cla}, this calculation was performed for a neutral fluid compactified on a torus.} From them, we read off the following expressions:
\begin{align}\label{eq:firstorder}
\begin{split}
\eta(x) &= e^{2\alpha\phi} \, \hat \eta (x)
\cosh \varphi (x) \, \, ,
\\
{\boldsymbol \kappa}_{m n}(x)
&=
 e^{2\alpha\phi} \, \hat\eta (x) \cT (x) \cosh \varphi(x)  \,  \left(
 	\delta_{m n}
 	- {\sinh^4 \varphi \over \cosh^2 \varphi} \, \cV_m{}^\alpha {\bf n}_\alpha \cV_n{}^\beta  {\bf n}_\beta 
 	\right)(x)
\, .
\\
\zeta(x)
&=
2  e^{2\alpha\phi(x)} \, \hat \eta (x) \, \cosh \varphi \left[
	\frac{1}{d-n}
	+ c_s^4
	\left(
		1 
		- 
		\cosh^4 \varphi {1 \over p} \frac{\partial \hat p}{\partial \heps}
		+ e^{2\alpha\phi(x)} \cosh^5 \varphi \, 
		\hat\zeta \, \left( \frac{\partial \hat p}{\partial \heps}\right)^2
	\right)
\right] .
\end{split}
\end{align}
It is important to stress that when getting rid of any dependence on the scalar fields $\varphi 
$, we recover the $d$-dimensional quantities multiplied by the dilaton factor $e^{2\alpha\phi}$, which parametrizes the volume of the internal manifold. On the other hand, it is worth to mention that the non-Abelian behavior of the conductivity matrix arises from the dependence of the scalar vielbein $\cV_m{}^{\alpha}$.

The analysis in this section demonstrates that the non-Abelian KK dimensional reduction is an {\sl ab initio} and efficient method for deriving the structure and dynamics of Yang-Mills plasma. Moreover, the construction that leads to Eq.~\eqref{eq:firstorder} gives a hydrodynamic frame-independent transport. We see from Eq.~\eqref{eq:firstorder} that, apart from viscosities, we have the non-Abelian conductivity matrix ${\boldsymbol \kappa} _{mn}$, which is directly connected to the non-Abelian degrees of freedom in the system. We remark that a similar quantity was obtained in the context of the fluid/gravity duality \cite{Eling:2010hu}.

Now that we have clearly formulated non-Abelian hydrodynamics, we can study various related issues. Understanding conductivity is a major challenge in recent approaches to holographic superfluids. One can show that, at the phase transition, a set of SU(2) currents can be used as an order parameter \cite{Gubser:2008wv}. Moreover, it was observed in \cite{Herzog:2014tpa} that employing a non-Abelian gauge transformation allows one to obtain a finite conductivity without breaking translational symmetry.

On the other hand, this theory results a very suitable and robust framework where to study the quark-gluon plasma. In this respect, one important phenomenon of this system is the study of the relaxation time. This is the time at which the non-Abelian character of the plasma is relaxed, thus becoming purely Abelian. This is a known property that has not been theoretically understood neither for quark-gluon plasma nor for spintronics systems.\footnote{It is worth to mention that our system can be coupled to additional fermionic degrees of freedom, as we are using the vielbein formalism.} Since our construction can describe the dissipative part of non-Abelian hydrodynamics, we expect it to be useful in elucidating the relaxation mechanism of the color current.


\section{Outlooks}

In this work, we have proposed a new approach for constructing non-Abelian hydrodynamics, consisting of colored fluid interacting with Yang-Mills theory. Based on non-Abelian KK dimensional reduction, the simplicity, clarity, and elegance of the proposed formulation enables one to understand the properties of Yang-Mills plasma.

We have presented an {\sl ab initio} approach for constructing hydrodynamics charged under both Maxwell and Yang-Mills plasma. With the non-Abelian KK reduction, we have compactified the Einstein-fluid equations on a group manifold. We have started with the most general dissipative, neutral fluid. After the reduction, we have obtained Yang-Mills plasma equations for a dissipative, colored fluid interacting  non-Abelian gauge fields.  Though having done the reduction on $\mathbb{S}^1$ and SU(2) group manifold, this procedure can be applied to any type of group manifold. Our approach is not restricted by symmetries that are only symmetries of the Lagrangian. Hence, the KK reduction approach seems to be a robust and covariant method to naturally obtain hydrodynamics coupled to (non-)Abelian gauge fields. The method straightforwardly extends to dissipative hydrodynamics coupled to gravity and a specific form of dilaton scalar field, which would also bear applications to early universe cosmology, formation of large-scale structure or compact objects, and colored turbulence.

We have studied the conservation laws of colored fluid and obtained a non-Abelian covariantly conserved current ${\bf J}_{am}$, which is proportional to the fluid velocity field, as predicted by \cite{Jackiw:2004nm}. In addition, truncating the scalar fields coming from the gravity sector to be constant, we obtain the equation for non-Abelian Lorentz force.

We have shown that the reduction procedure does not preserve the hydrodynamic frames. As a consequence, the effective transport coefficients cannot be straightforwardly read off from the reduced system. We have proposed a frame-independent formulation of dissipative fluids for the non-Abelian gauge fields that is thermodynamically valid and generalizes the one given in \cite{Bhattacharya:2011tra}. With this construction, we identified the $d$-dimensional dissipative magnitudes that characterize the effective fluid in terms of the $D$-dimensional ones. In particular, we have obtained a conductivity matrix whose non-Abelian nature is given by the scalar vielbein $\cV_m{}^\alpha$.


The Yang-Mills plasma equations are in complete agreement with the equations of Maxwell plasma derived in Section \ref{sec:u1}. If we set the structure constants $f_{mn}{}^p=0$, we can check that these equations are reduced to the equations for charged fluid coupled to $U(1)^3$ Abelian gauge fields. The results of this section can also be straightforwardly extended to other, higher-dimensional group manifold $G$. We claim that, for fixed $d$, the large-$D$ limit should be taken seriously as it corresponds to the limit for which rank$(G)$ gets large, revealing a new perspective to the planar limit of Yang-Mills plasma. Results on this aspect will be relegated to a separate publication.

We believe the proposed approach marks an important advance toward the understanding of the evolution of nuclear matter after a heavy-ion collision. Hydrodynamics with non-Abelian degrees of freedom that have not thermalized is a transient phase and the lack of a first-principle derivation of the equations that govern its evolution has been a major obstacle for further developments.

Having now the  \emph{ab initio} construction of fluid and field equations, we can utilize complementary methods such as kinetic theory or gauge$/$gravity duality to shed more light of this regime. Gravitational solutions with Abelian gauge fields have recently been studied \cite{Compere:2011dx, Kanitscheider:2008kd,Liu:2016njg}. Therefore, we provide a robust formulation of non-Abelian hydrodynamics where to test fluid/gravity duality beyond Abelian fluids.

In addition to a phenomenological description of quark-gluon plasma, recent formulation of fluid dynamics in terms of fluid/gravity duality has increased the interest in the analysis of fluids coupled to Yang-Mills fields. In this picture, fluid is a field theory dual to a black hole in higher-dimensional, asymptotically anti-de Sitter spacetime (see \cite{Rangamani:2009xk} for a review). It would be interesting to further explore the physics of black holes with non-Abelian and dilatonic hairs using the non-Abelian Kaluza-Klein reduction \cite{Gouteraux:2011qh}.


\section*{Acknowledgements}
We thank Yong-Min Cho, Richard Davison, Sa\v{s}o Grozdanov, Seungho Gwak, Dima Kharzeev, Jaewon Kim, Kanghoon Lee, Andy Lucas, Jeong-Hyuck Park, Malcolm Perry, Chris Pope, Woohyun Rim, Alejandro Rosabal, Yuho Sakatani, and Jan Zaanen for useful discussions. SJR and PS acknowledge hospitality of NORDITA program ``{\sl Holography and Dualities 2016}" during the final stage.
The work of  JJFM was supported by the Fundaci\'on S\'eneca - Talento Investigador Program. The work of SJR was supported in part by the National Research Foundation Grants 2005-0093843, 2010-220-C00003 and 2012K2A1A9055280. SJR was also supported in part by the Munich Institute for Astro- and Particle Physics (MIAPP) of the DFG cluster of excellence "Origin and Structure of the Universe".
The work of PS was supported by a Marie Curie International Outgoing Fellowship, grant number PIOF-GA-2011-300528.

\appendix

\section{Einstein equations on a group manifold}
\label{app:su2details}
In this appendix, we elaborate technical details of the non-Abelian Kaluza-Klein compactification on a group manifold. We also explain the convention used in this work.

We will consider that our starting system is defined on a $D$ dimensional manifold ${\cal M}_D(\hat g)$ with coordinates $\hat x^M$, for $M=1,\ldots,D$. For the tangent spacetime description we introduce a vielbein $E_M{}^A$, where $A=1,\ldots,D$, which satisfies
\begin{equation}
\hat g_{MN}(\hat x) = E_M{}^A(\hat x) E_N{}^B(\hat x) \eta_{AB}
\,, \qquad \qquad \eta_{AB} = (-+\ldots+) \, .
\end{equation}

\subsection{General ansatz}
We will perform a KK dimensional reduction. To do so, we will assume that ${\cal M}_D(\hat g)= {\cal M}_d(g) \times X_n({\boldsymbol M})$. ${\cal M}_d(\hat g)$ is the $d$-dimensional external spacetime manifold on which our resulting system will live whereas $X_n({\boldsymbol M})$ is the $n$-dimensional internal manifold. The coordinates are split as $\hat x^M=\{x^\mu, y^m \}$, where $\mu=1,\ldots,d$ and $m=1,\ldots,n$. Despite the scalar matrix $M_{m n}$ will parametrize the fluctuations of the internal manifold, the final $d$-dimensional system cannot have any functional dependence on $X_n$. The

We start with the reduction ansatz for the vielbein expressed in terms of the Maurer-Cartan one-forms:
\begin{align}
\begin{split}
\hat E^a (x, y)
&=
e^{\alpha\phi(x)} e^a(x)
\, ,
\\
\hat E^\alpha(x, y)
&=
g^{-1} e^{\beta\phi(x)} \cV_m{}^\alpha(x) ({\boldsymbol \sigma}^m - g{\bf A}^m(x))
\, ,
\end{split}
\end{align}
where ${\boldsymbol \sigma}^m  \equiv \mathfrak u_n{}^m(y) \rmd y^n$ are the twist matrices, which will depend on the group manifold coordinates $y$. Here, $g$ is a gauge coupling parameter.

We will compute various geometric quantities. The spin-connection is defined as
\begin{align}
\hat\omega_{C,AB} &= -\hat\Omega_{CA,B}+ \hat\Omega_{AB,C} - \hat\Omega_{BC,A}
\,
\end{align}
where
\begin{align}
\hat\Omega_{AB,C} &= \frac12\left( E_A{}^M E_{B}^N - E_B{}^M E_{A}{}^N \right) \partial_N E_{M}{}^D \hat\eta_{D C}
\, .
\end{align}
Substituting the vielbein ansatz, 
we obtain the following expressions:
\begin{align}
\begin{split}
\hat\omega_{c, ab} &= e^{-\alpha\phi} \left[ \omega_{c,ab} + 2\alpha \eta_{c[a} \partial_{b]}\phi \right]
\, ,
\\
\hat\omega_{c, a\beta} &=
-\frac{1}{2} e^{(-2\alpha+\beta)\phi} {\bf F}_{\mu\nu}{}^n e_c{}^\mu e_a{}^\nu \cV_{n\beta}
\, , \\
\hat\omega_{c, \alpha\beta} &=
+\frac12 e^{-(\alpha+\beta)\phi} \left[
	\cV_\alpha{}^m e_c{}^\mu D_\mu(e^{\beta\phi} \cV_{m \beta})
	- \cV_\beta{}^m e_c{}^\mu D_\mu(e^{\beta\phi} \cV_{m \alpha})
	\right]
\, , \\
\hat\omega_{\gamma, a b}
&=
+ \frac12 e^{(-2\alpha+\beta)\phi} {\bf F}_{\mu\nu}{}^m e_a{}^\mu e_b{}^\nu \cV_{m\gamma}
\, , \\
\hat\omega_{\gamma, a \beta}
&=
-\frac12 e^{-(\alpha+\beta)\phi} e_a{}^\mu \cV_\beta{}^m \cV_\gamma{}^n D_\mu(e^{\beta\phi} {\boldsymbol M_{mn}} )
\, , \\
\hat\omega_{\gamma, \alpha \beta}
&=
+\frac g2 e^{-\beta\phi} f_{m n}{}^p \left[
	\cV_{\gamma}{}^n \cV_\alpha{}^m \cV_{p\beta} + \cV_\alpha{}^m \cV_\beta{}^n \cV_{p\gamma} -\cV_\beta{}^m \cV_\gamma{}^n \cV_{p \alpha}
	\right]
\, ,
\end{split}
\end{align}
where
\begin{align}
\begin{split}
{\boldsymbol M_{mn}}
&=
\cV_m{}^\alpha \cV_n{}^\beta \delta_{\alpha\beta}
\, , \\
{\bf F}^m{}_{\mu\nu}
&\equiv
\partial_{\mu} {\bf A}^m{}_{\nu} - \partial_\nu {\bf A}^m_\mu  +  g f_{n p}{}^m {\bf A}^n{}_{\mu} {\bf A}^p{}_\nu
\, , \\
D_\mu \cV_{m}{}^\alpha
&\equiv
\partial_\mu \cV_{m}{}^\alpha - g f_{n m}{}^p {\bf A}^n{}_{\mu} \cV_{p}{}^\alpha
\,
\end{split}
\end{align}
and
\begin{align}
f_{mn}{}^p = - (\mathfrak u^{-1})_m{}^s (\mathfrak u^{-1})_n{}^t (\partial_{s} \mathfrak u_{t}{}^p - \partial_{t} \mathfrak u_{s}{}^p)
\, .
\end{align}

We will calculate the components of the Ricci tensor $\hat R_{AB}= \hat{R}_{ACBD} \hat \eta^{CD}$ and the scalar curvature $\hat R=\hat R_{A B} \hat \eta^{A B}$ by substituting the components of the spin connection $\hat \omega_{CAB}$ into the expression for the Riemann tensor,
\begin{align}
\hat R_{ACBD}
&=
\partial_A\hat \omega_{CBD} - \partial_C\hat \omega_{ABD}
+ \hat \omega_{A B}{}^E \hat\omega_{C E D} - \hat \omega_{CB}{}^E \hat\omega_{A E D}
\, .
\end{align}

\subsection{SU(2) group manifold}
In what follows, we restrict to the SU(2) group manifold, so that $f_{mn}{}^p$ will be the SU(2) structure constants, $f_{mnp} = \epsilon_{mnp}$. In this case, the components of the spin connection are given by \cite{Lu:2002uw}
\begin{align}
\begin{split}
\hat \omega_{ab}
&=
\omega_{ab}
+ 2\alpha e^{-\alpha \phi} \eta_{c[a} D_{b]}\phi \hat e^c
+  \frac12 e^{(-2\alpha+\beta)\phi} {\bf F}^\beta{}_{ab} \hat e^\beta
\, , \\
\hat \omega_{a\beta}
&=
-e^{-\alpha \phi}  \mathbb P_{a \beta\gamma}  \hat e^\gamma
- \beta e^{-\alpha \phi} D_a\phi \hat e^\beta
+ \frac12 e^{(-2\alpha+\beta)\phi}  {\bf F}^\beta{}_{ab}  \hat e^b
\, , \\
\hat\omega_{\alpha \beta}
&=
e^{-\alpha \phi} \mathbb Q_{a \alpha\beta } \hat e^a
+ \frac g2 e^{-\beta\phi} ( \mathbb M^{\gamma \delta} \epsilon_{\alpha \beta \delta} + \mathbb M^{\beta \delta}\, \epsilon_{\alpha\gamma \delta}
- \mathbb M^{\alpha\delta} \epsilon_{\beta\gamma\delta})  \hat e^\gamma
\, .
\end{split}
\end{align}
where
\begin{align}
{\bf F}^\alpha{}_{ab}
&\equiv
\cV_m{}^\alpha {\bf F}^m{}_{ab}
\, ,
\end{align}
$M_{\alpha\beta}$ is the SU(2) covariant scalar matrix
\begin{align}
\mathbb M^{\alpha \beta}
&\equiv
\cV_m{}^\alpha \cV_n{}^\beta {\delta}^{mn}
\, ,
\end{align}
and
\begin{align}
\mathbb P_{a\, \alpha\beta} &\equiv \frac12 [ \cV_\alpha{}^m\, D_a \cV_m{}^\beta +
                           \cV_\beta{}^m\, D_a \cV_m{}^\alpha]\,,
&
\mathbb Q_{a\, \alpha\beta} &\equiv \frac12 [ \cV_\alpha{}^m\, D_a \cV_m{}^\beta -
                           \cV_\beta{}^m\, D_a \cV_m{}^\alpha]\, .
\end{align}

The Ricci tensor components are
\begin{align}
\begin{split}
\hat R_{ab}
=&
e^{-2\alpha \phi} \left[
	R_{ab}
	- \tfrac12 \partial_a \phi \partial_b \phi - \mathbb P_{a \, \alpha\beta}  \mathbb P_{b \, \alpha \beta}
	- \alpha \Box \phi  \eta_{ab}
	- \tfrac12 e^{-\frac23 \alpha (n+1) \phi}  \mathbf F ^ \alpha{}_{a c} \mathbf F^ \alpha{}_{b d}  \eta^{cd}
	\right]
\, ,
\\
\hat R_{a \beta }
=&
-\tfrac12 e^{\frac 1 3 \alpha  (n-5) \phi } \left[
	\cD^b( e^{-\frac 2 3 \alpha(n+1) \phi} \mathbf F^\beta{}_{ab})
	+ e^{-\frac23 \alpha (n+1)\phi}  \mathbf F^\gamma{}_{ab}  \mathbb P^b{}_{\beta \gamma}
	- 2 g \epsilon_{\beta\gamma \delta}  \mathbb M^{\delta\lambda} \mathbb P_{a  \gamma \lambda}
	\right]
\, ,
\\
\hat R_{\alpha\beta}
=&
-\tfrac12 e^{-2\alpha \phi}\left[
	\cD_a \mathbb P^a{}_{\alpha\beta}
	- \tfrac23 \alpha (d-2) \Box \phi \delta_{\alpha\beta}
	- \tfrac12 e^{-\frac23 \alpha (d+1) \phi} \mathbf F^\alpha{}_{ab} \mathbf F^\beta{}_{cd} \eta^{ac} \eta^{bd}
\right.
\\
&
\left.
\qquad
	- 4 g^2 e^{\frac23\alpha (d+1) \phi}\left(
		\mathbb M^{\alpha\gamma} \mathbb M^{\beta \gamma}
		-\tfrac12 \mathbb M^{\alpha\beta} \mathbb M
		\right)
	+ 2 g^2 e^{\frac23\alpha (d+1) \phi}\left(
		\mathbb M^{\gamma\delta} \mathbb M^{\gamma \delta}
 		-\tfrac12 \mathbb M^2
 		\right) \delta_{\alpha\beta}
 	\right]
\, ,
\end{split}
\end{align}
where
\begin{align}
\mathbb M\equiv \mathbb M_{\alpha\alpha} \qquad \mbox{and} \qquad \cD_a \cV_m{}^\alpha = D_a \cV_m{}^\alpha + \mathbb Q_{a \alpha\beta} \cV_m{}^\beta
\, .
\end{align}

\subsection{Equations of motion}
Our starting point is the $D$-dimensional Einstein-fluid equation
\begin{align}
{\cal G}_{MN} \equiv \hat R_{M N} - \left( \hat T_{MN} + \frac{1}{D-2}g_{MN} \hat T  \right)
=
0
\, .
\end{align}
We will analyze the tensor ${\cal G}_{AB} = E_A{}^M E_B{}^N {\cal G}_{MN}$:
\begin{align}
{\cal G}_{AB}\equiv \hat R_{AB} - \left( \hat T_{AB} + \frac{1}{D-2}\eta_{AB} \hat T  \right)
=0
\, .
\end{align}
Here, we analyze each components of the Einstein-fluid equation.  We begin with the internal components, ${\cal G}_{m n}$:
\begin{align}
{\cal G}_{mn} = g^{-2} e^{2\beta\phi} \mathfrak u_{m}{}^p \mathfrak u_n{}^q \cV_p{}^\alpha \cV_q{}^\beta {\cal G}_{\alpha\beta}
\, .
\end{align}
On one hand, this equation has to be satisfied for any scalar fields $\cV_m{}^\alpha$. As the twist matrices $\mathfrak u_m{}^n$ depend on internal coordinates, we have that ${\cal G}_{\alpha\beta}=0$, where ${\cal G}_{\alpha\beta}$ is given by
\begin{dmath}
{\cal G}_{\alpha\beta}
=
-\tfrac12 e^{-2\alpha \phi}\left[
	\cD_a  \mathbb P^a{}_{\alpha\beta}
	- \tfrac23 \alpha (d-2) \Box \phi \delta_{\alpha\beta}
	- \tfrac12 e^{-\frac23 \alpha (d+1) \phi} \mathbf F^\alpha{}_{ab} \mathbf F^\beta{}_{cd} \eta^{ac} \eta^{bd}
	- 4 g^2 e^{\frac23\alpha (d+1) \phi}\left(
		\mathbb M^{\alpha\gamma} \mathbb M^{\beta \gamma}
		-\tfrac12 \mathbb M^{\alpha\beta} \mathbb M
		\right)
	+ 2 g^2 e^{\frac23\alpha (d+1) \phi}\left(
		\mathbb M^{\gamma\delta} \mathbb M^{\gamma \delta}
 		-\tfrac12 \mathbb M^2
 		\right) \delta_{\alpha\beta}
 	\right]
+ \frac{1}{d+1}\delta_{\alpha\beta} \hat T^{\rm fluid} - \hat T_{\alpha\beta}^{\rm fluid}
\, ,	
\label{eq:raw-EOMV}
\end{dmath}
Solving the trace part, ${\cal G}_{\alpha\beta}\delta^{\alpha\beta}=0$, where
\begin{dmath}
{\cal G}_{\alpha\beta} \delta^{\alpha\beta}
= e^{-2\alpha\phi} (d-2)\left[
\alpha \Box \phi
+ \tfrac{1}{4(d-2)} e^{-\frac{2}{3}\alpha(d+1)\phi} (\mathbf F^{\alpha})^2
- \tfrac{g^2}{d-2} e^{\frac{2}{3}\alpha(d+1)\phi} (\mathbb M^{\alpha\beta} \mathbb M^{\alpha\beta} - \tfrac12 \mathbb M^2)
+ \tfrac{1}{d-2} e^{2\alpha\phi}(
	\tfrac{3}{d+1} \hat T^\fluid - \hat T^\fluid_{\alpha\beta} \delta^{\alpha\beta}
	)
	\right]
\, ,
\end{dmath}
we solve for $\Box\phi$ and substitute back to Eq.\eqref{eq:raw-EOMV}. We then obtain from ${\cal G}_{\alpha\beta}=0$ that
\begin{align}
\cD^a \mathbb P_{a \alpha\beta}
&=
\tfrac12 e^{-\frac23\alpha(d+1)\phi}\left[
	\mathbf F^\alpha{}_{ab} \mathbf F^{\beta}{}_{cd} \eta^{ac} \eta_{bd}
	- \tfrac13 (\mathbf F^\gamma)^2 \delta_{\alpha\beta}
	\right]
\nonumber\\
&
+4g^2 e^{\frac23\alpha(d+1)\phi}\left[
	\mathbb M^{\alpha\gamma} \mathbb M^{\beta\gamma}
	-\tfrac12 \mathbb M^{\alpha\beta} \mathbb M
	-\tfrac13 \left(
		\mathbb M^{\gamma\delta} \mathbb M^{\gamma \delta}
		-\tfrac12 \mathbb M^2
		\right) \delta_{\alpha\beta}
	\right]
\nonumber\\
&
+ 2 e^{2\alpha\phi}\left[
	\tfrac1{d+1} \hat T^{\rm fluid} \eta_{\alpha\beta}
	- \hat T_{\alpha\beta}^{\rm fluid}
	- \tfrac13\left(
		\tfrac{3}{d+1}\hat T^{\rm fluid}
		- \hat T^{\rm fluid}_{\gamma\delta} \delta^{\gamma\delta}
		\right) \delta_{\alpha\beta}
	\right]
\, .
\end{align}

Let us consider now the field equations ${\cal G}_{\mu n}= 0$. As
\begin{align}
{\cal G}_{\mu n}
= E_\mu{}^A E_n{}^\beta {\cal G}_{A \beta} = E_\mu{}^a E_n{}^\beta {\cal G}_{a \beta} 
=
g^{-1} e^{(\alpha+\beta)\phi} \mathfrak u_m{}^p e_\mu{}^a \cV_p{}^\beta {\cal G}_{a\beta}
\, ,
\end{align}
it follows that ${\cal G}_{\mu n} = 0$ is equivalent to ${\cal G}_{a\beta} = 0$.

We now study the field equations ${\cal G}_{a \beta}=0$.  We have 
\begin{dmath}
\cD^b(e^{-\frac23\alpha(d+1)\phi} \mathbf F^\beta{}_{ab})
+e^{-\frac23\alpha(d+1)\phi} \mathbf F^{\gamma}{}_{a b} \mathbb P^b{}_{\beta \gamma}
-2g \epsilon_{\beta\gamma\delta} \mathbb M^{\delta\lambda} \mathbb P_{a \gamma \lambda}
+ 2  e^{-\frac13\alpha(d-5)\phi} \hat T^{\rm fluid}_{a \beta}
= 0
\, .
\label{eq:EOMA-su2}
\end{dmath}

Finally, let us consider the $d$-dimensional components of the Einstein field equations, ${\cal G}_{\mu\nu}= 0$. Using the equations for other components, this equation implies
\begin{align}
{\cal G}_{\mu\nu} = E_\mu{}^A E_\nu{}^B {\cal G}_{AB} = e^{2\alpha\phi} e_\mu{}^a e_\nu{}^b {\cal G}_{ab} = 0
\, .
\end{align}
Therefore the resulting equation is ${\cal G}_{ab} = 0$, where
\begin{dmath}
{\cal G}_{ab}
=
R_{ab}
-\tfrac12 \partial_a\phi\partial_b\phi
- \mathbb P_{a \beta\gamma} \mathbb P_{b \beta\gamma}
-\tfrac12 e^{-\frac23\alpha(d+1)\phi}\left[
	\mathbf F^\gamma{}_{a c} \mathbf F^\gamma{}_{b d} \eta^{cd}
	- \tfrac{1}{2(d-2)}(\mathbf F^\gamma)^2 \eta_{ab}
	\right]
-\tfrac{g^2}{d-2} e^{\frac23\alpha(d+1)\phi} (\mathbb M^{\gamma\delta} \mathbb M^{\gamma\delta} - \tfrac12 \mathbb M^2) \eta_{ab}
- e^{2\alpha\phi}\left[
	\hat T^{\rm fluid}_{ab}
	-\tfrac{1}{d-2} \hat T^{\rm fluid}_{cd} \eta^{cd} \eta_{ab}
	\right]
\, .
\end{dmath}
These are the Einstein equations for the $d$-dimensional system, which can be equivalently rewritten as
\begin{align}
R_{\mu\nu} - \frac{1}{2}g_{\mu\nu} R = T^{\tot}_{\mu\nu}
\, ,
\end{align}
where $T^{\tot}_{\mu\nu}= e_\mu{}^a e_\nu{}^b T^{\tot}_{ab}$ is
\begin{dmath}
T^{\tot}_{ab}
=
 \frac12 \left(\partial_{a}\phi \partial_b\phi - \frac12 (\partial\phi)^2\eta_{ab}\right)
+ \mathbb P_{a\beta\gamma} \mathbb P_{b \beta\gamma} - \frac12 \mathbb P^2 \eta_{ab}
+ e^{-\frac23\alpha(d+1)\phi}\left[ \frac12
	\mathbf F^\alpha{}_{a c} \mathbf F^\alpha{}_{b d} \eta^{cd}
	- \frac14 (\mathbf F^\gamma)^2 \eta_{ab}
	\right]
-g^2 e^{\frac23\alpha(d+1)\phi}\left(
	\mathbb M^{\gamma\delta} \mathbb M^{\gamma\delta}
	- \frac12 \mathbb M^2
	\right)\eta_{ab}
	 +e^{2\alpha\phi} \hat T^{\rm fluid}_{ab}
\, .
\end{dmath}
From this expression, we also see that the energy-momentum tensor of non-Abelian hydrodynamics
$T^{\rm fluid}_{ab}$ is given by
\begin{align}
T^{\rm fluid}_{ab} =  e^{2\alpha\phi} \hat T^{\rm fluid}_{ab}
\,.
\label{eq:effectiveTsu2app}
\end{align}

\section{Conservation laws}
In this section we will calculate the conservation laws of the $d$-dimensional theory, namely the current conservation and the Lorentz force. Despite of not making any assumption on the scalar fields, after obtaining the most general expressions we will study the cases for which scalar fields are covariantly constant,
\begin{align*}
D_\mu \cV_m{}^a= D_\mu \phi = 0
\, ,
\end{align*}
in order to make contact with the conservation laws considered in hydrodynamics, where no  degrees of freedom associated to scalar fields take place.
\subsection{Current conservation}
Current conservation follows from the consistency condition of the EOMs.
Before applying a covariant derivative $\cD_a$ on  \eqref{eq:EOMA-su2}, we first rewrite the equation of motion for gauge field as
\begin{align}
\left[
	D^b\left(
		e^{-\frac23\alpha(d+1)\phi} \mathbb M_{nm} \mathbf F^m{}_{ab}
		\right)
	-2g \epsilon_{\sigma\gamma\delta} \mathbb M^{\delta\lambda} \mathbb P_{a \gamma \lambda} \cV_n{}^\sigma
	+ 2  e^{-\frac13\alpha(d-5)\phi} \hat T^{\rm fluid}_{a \gamma} \cV_n{}^\gamma
\right] \cV_\beta{}^n
= 0
\, .
\end{align}
As ${\cal V}_\beta{}^n$ is non-degenerate in general, without loss of generality, we can assume the equation of motion to be
\begin{align}
	D^b\left(
		e^{-\frac23\alpha(d+1)\phi} \mathbb M_{nm} \mathbf F^m{}_{ab}
		\right)
	-2g \epsilon_{\sigma\gamma\delta} \mathbb M^{\delta\lambda} \mathbb P_{a \gamma \lambda} \cV_n{}^\sigma
	+ 2  e^{-\frac13\alpha(d-5)\phi} \hat T^{\rm fluid}_{a \gamma} \cV_n{}^\gamma
= 0
\, .
\end{align}
Next, taking the divergence of covariant derivative $D_a$, we obtain
\begin{align}
D^a\left(
	2g \epsilon_{\sigma\gamma\delta} \mathbb M^{\delta\lambda} \mathbb P_{a \gamma \lambda} \cV_n{}^\sigma
	- 2 \epsilon e^{-\frac13\alpha(d-5)\phi} \hat T^{\rm fluid}_{a \gamma} \cV_n{}^\gamma
\right)
= 0
\, .
\label{eq:currentconservapp}
\end{align}
This allows us to define color conserved current
\begin{align}
{\bf J}_{m a}
& =
	2g \epsilon_{\sigma\gamma\delta} \mathbb M^{\delta\lambda} \mathbb P_{a \gamma \lambda} \cV_m{}^\sigma
	- 2 \epsilon e^{-\frac13\alpha(d-5)\phi} \hat T^{\rm fluid}_{a \gamma} \cV_m{}^\gamma
	\, .
\label{eq:currentsu2}
\end{align}

If we set the scalar fields  $D_a \cV_m{}^\beta=0$, then $\mathbb P_{a\beta\gamma}=0$ and the color current will be purely associated to the off-diagonal components of the $D$-dimensional fluid energy-momentum tensor.

\subsection{Lorentz force}
To study the Lorentz force, we will make use of the Bianchi identity of the Einstein tensor
\begin{align}
\nabla^\mu \left( R_{\mu\nu} - \tfrac12 g_{\mu\nu}R\right) = \nabla^\mu T^\tot_{\mu\nu} = 0
\, .
\end{align}
Upon vielbein compatibility, this is equivalent to
\begin{align}
\mathfrak{D}^aT^\tot_{ab}=0
\, ,
\end{align}
where $\mathfrak{D}_a = e_a{}^\mu (\partial_\mu + \omega_\mu)$, where $\omega$ is the $d$-dimensional spin connection.
Explicitly,
\begin{dmath}
\mathfrak{D}^aT^\tot_{ab}
=
\mathfrak{D}^a\left(
	e^{2\alpha\phi}\hat T^{\rm fluid}_{ab}
	\right)
+\tfrac12\left(
	D^a\partial_a\phi \partial_b \phi
	+\partial_a\phi D^a\partial_b \phi
	-\partial_c\phi D^a\partial_c \phi\eta_{ab}
	\right)
+\cD^a \mathbb P_{a\beta\gamma}   \mathbb P_{b\beta\gamma}
+\mathbb P_{a\beta\gamma}\cD^a \mathbb P_{b\beta\gamma}-\cD^a \mathbb P_{c\beta\gamma} \mathbb P_{c\beta\gamma}\eta_{ab}
+\tfrac12 \mathfrak{D}^a\left( e^{\frac23\alpha(d+1)\phi}\left[
	\mathbf F^\alpha{}_{a c} \mathbf F^\alpha{}_{b d} \eta^{cd}
	- \tfrac12 (\mathbf F^\gamma)^2 \eta_{ab}
	\right]
	\right)
+\tfrac12e^{\frac23\alpha(d+1)\phi}\cD^a\left[
	\mathbf F^\alpha{}_{a c} \mathbf F^\alpha{}_{b d} \eta^{cd}
	- \tfrac12 (\mathbf F^\gamma)^2 \eta_{ab}
	\right]
-g^2 \mathfrak{D}^a\left[
	e^{\frac23\alpha(d+1)\phi}\left(
		\mathbb M^{\gamma\delta} \mathbb M^{\gamma\delta}
		- \tfrac12 \mathbb M^2
		\right)\eta_{ab}
	\right]
\, .
\end{dmath}
Let us analyze various terms separately.
\begin{dmath}
\tfrac12\left(
	\mathfrak{D}^a\partial_a\phi \partial_b \phi
	+\partial_a\phi D^a\partial_b \phi
	-\partial_c\phi D^a\partial_c \phi\eta_{ab}
	\right)
=
\alpha\partial_b\phi\left[
	-\tfrac{1}{8(d-2)} e^{-\frac23\alpha(d+1)\phi}(\mathbf F^\gamma)^2
	+\tfrac1{2(d-2)} g^2 e^{\frac23\alpha(d+1)\phi}
	-\tfrac{1}{2(d-2)} e^{2\alpha\phi}\left(
		\tfrac3{d+1}\hat T
		-\hat T_{\lambda\sigma} \delta^{\lambda\sigma}
		\right)
	\right]
\, ,
\end{dmath}
and
\begin{dmath}
	\cD^a \mathbb P_{a\beta\gamma}  \mathbb P_{b\beta\gamma}
	+ \tfrac12 \cD_{[a} \mathbb P_{b]\beta\gamma} \mathbb P_{a\beta\gamma}
=
\tfrac12e^{-\frac23\alpha(d+1)\phi}\left[
	\mathbf F^\beta \mathbf F^\gamma - \tfrac13(\mathbf F^\sigma)^2 \delta_{\beta\gamma}
	\right]\mathbb P_{b\beta\gamma}
+4 g^2 e^{\frac23\alpha(d+1)\phi}\left[
	\mathbb M^{\beta\lambda} \mathbb M^{\gamma\lambda}
	- \tfrac12 \mathbb M^{\beta\gamma} \mathbb M
	- \tfrac13\delta_{\beta\gamma}V(\mathbb M)
	\right] \mathbb P_{b\beta\gamma}
- 2 \mathbb P_{[a|\beta\lambda} \cV_\lambda{}^m D_{|b]}\cV_m{}^\gamma \mathbb P_{a\beta\gamma}
- \frac{g}{2} \cV_\beta{}^m f_{nm}{}^p \mathbf F^n{}_{ab} \cV_p{}^\gamma \mathbb P_{a\beta\gamma} \, .
\end{dmath}
Using the Bianchi identity $D \mathbf F^m=0$ and the above equations of motion, we have
\begin{dmath}
\tfrac12\cD^a\left\{e^{-\frac23\alpha(d+1)\phi}\left[
	\mathbf F^\alpha{}_{a c} \mathbf F^\alpha{}_{b d} \eta^{cd}
	- \tfrac12 (\mathbf F^\gamma)^2 \eta_{ab}
	\right]
	\right\}
=
\tfrac12\cD^a\left(e^{-\frac23\alpha(d+1)\phi} \mathbb M_{mn} \right)\left[
	\mathbf F^m{}_{a c} \mathbf F^n{}_{b d} \eta^{cd}
	- \tfrac12 \mathbf F^m \mathbf F^n \eta_{ab}
	\right]
-\tfrac12\cD^a\left(e^{-\frac23\alpha(d+1)\phi}\right) \mathbf F^\alpha{}_{a c} \mathbf F^\alpha{}_{b c}
+ e^{-\frac23\alpha(d+1)\phi} \mathbf F^\alpha{}_{cd} \mathbb F^\beta{}_{bc} \mathbb P^d{}_{\alpha\beta}
- g\epsilon_{\alpha\gamma\delta} \mathbb M^{\delta\lambda} \mathbb P_{c\gamma\lambda}\mathbf F^\alpha{}_{bc}
+2  e^{-\frac13\alpha(d-5)\phi} \hat T_{c\alpha} \cV_n{}^\alpha \mathbf F^n{}_{bc}
\, .
\end{dmath}
Summing up all the terms, we have
\begin{dmath}
\mathfrak{D}^aT^\tot_{ab}
=
\mathfrak{D}^a T^{\rm fluid}_{ab}
+
\alpha\partial_b\phi\left[
	-\tfrac{1}{8(d-2)} e^{-\frac23\alpha(d+1)\phi}( \mathbf F^\gamma)^2
	+\tfrac1{2(d-2)} g^2 e^{\frac23\alpha(d+1)\phi}
	-\tfrac{1}{2(d-2)} e^{2\alpha\phi}\left(
		\tfrac3{d+1}\hat T^{\rm fluid}
		-\hat T^{\rm fluid}_{\lambda\sigma} \delta^{\lambda\sigma}
		\right)
	\right]
+\tfrac12e^{-\frac23\alpha(d+1)\phi}\left[
	\mathbf F^\beta \mathbf F^\gamma - \tfrac13(\mathbf F^\sigma)^2 \delta_{\beta\gamma}
	\right] \mathbb P_{b\beta\gamma}
+4 g^2 e^{\frac23\alpha(d+1)\phi}\left[
	\mathbb M^{\beta\lambda} \mathbb M^{\gamma\lambda}
	- \tfrac12 \mathbb M^{\beta\gamma}\mathbb M
	- \tfrac13\delta_{\beta\gamma}V
	\right] \mathbb P_{b\beta\gamma}
- 2 \mathbb P_{[a|\beta\lambda} \cV_\lambda{}^m D_{|b]}\cV_m{}^\gamma \mathbb P_{a\beta\gamma}
- \frac{g}{2} \cV_\beta{}^m f_{nm}{}^p \mathbf F^n{}_{ab} \cV_p{}^\gamma \mathbb P_{a\beta\gamma}
+\tfrac12\cD^a\left(e^{-\frac23\alpha(d+1)\phi} \mathbb M_{mn} \right)\left[
	\mathbf F^m{}_{a c} \mathbf F^n{}_{b d} \eta^{cd}
	- \tfrac12 \mathbf F^m \mathbf F^n \eta_{ab}
	\right]
-\tfrac12\cD^a\left(e^{-\frac23\alpha(d+1)\phi}\right) \mathbf F^\alpha{}_{a c} \mathbf F^\alpha{}_{b c}
+ e^{-\frac23\alpha(d+1)\phi} \mathbf F^\alpha{}_{cd} \mathbf F^\beta{}_{bc} \mathbb P^d{}_{\alpha\beta}
- g\epsilon_{\alpha\gamma\delta} \mathbb M^{\delta\lambda} \mathbb P_{c\gamma\lambda} \mathbf F^\alpha{}_{bc}
+2 e^{-\frac13\alpha(d-5)\phi} \hat T_{c\alpha} \cV_n{}^\alpha F^n{}_{bc}
-g^2 D^a\left[
	e^{\frac23\alpha(d+1)\phi} V(\mathbb M) 
		\eta_{ab}
	\right]
\, .
\end{dmath}
If we assume flat Minkowski in $d$ dimensions and that scalar fields are covariantly constant,
\begin{align}
D_a \cV_m{}^\alpha = D_{a} \phi= 0
\, ,
\end{align}
this expression reduces to
\begin{align}
D^a T^{\rm fluid}_{ab}
+2 e^{-\frac13\alpha(d-5)\phi} \hat T_{c\alpha} \cV_n{}^\alpha \mathbf F^n{}_{bc}
=
e^{2\alpha\phi}\left(
	D^a \hat T^{\rm fluid}_{ab}+
	2 e^{-\tfrac13\alpha(d+1)}\hat T^{\rm fluid}_{c\alpha} \mathbf F^\alpha{}_{bc}
	\right)
=0
\, .
\label{eq:lorentzsu2app}
\end{align}

\bibliography{nonabelhydro-bib}
\bibliographystyle{utphys}


\end{document}